%% file: belle2_alp.tex
\documentclass[
superscriptaddress,
reprint,
 amsmath,
 amssymb,
 prl
]{revtex4-1}
\usepackage[caption=false]{subfig}
\usepackage{xspace}
\usepackage{graphicx}
\usepackage{dcolumn}
\usepackage{bm}
\usepackage{color}
\usepackage[normalem]{ulem} 

\def \belletwo {Belle\,II\xspace}
\def \superkekb {SuperKEKB\xspace}

\def \intlumfullwitherror {$(496\pm 3)\,\text{pb}^{-1}$\xspace}
\def \intlum {$445\,\text{pb}^{-1}$\xspace}
\def \intlumwitherror {$(445\pm 3)\,\text{pb}^{-1}$\xspace}
\newcommand{\mamin}{\ensuremath{0.2\,\text{GeV/$c$$^2$}}\xspace}
\newcommand{\mamax}{\ensuremath{9.7\,\text{GeV/$c$$^2$}}\xspace}
\newcommand{\maminnu}{\ensuremath{0.2}\xspace}

\def \nfitsdiphoton {378\xspace}
\def \nfitsrecoil {124\xspace}
\newcommand{\fitrangetransition}{\ensuremath{6.85\,\text{GeV/$c^2$}}\xspace} 
\newcommand{\fitrangetransitionnu}{\ensuremath{6.85}\xspace} 

\newcommand{\mrecoilsq}{\ensuremath{M_{\text{recoil}}^2}\xspace}
\newcommand{\minvsq}{\ensuremath{M_{\gamma\gamma}^2}\xspace}

\newcommand{\fitrangeminnu}{\ensuremath{0.2}\xspace} 
\newcommand{\fitrangemax}{\ensuremath{9.7\,\text{GeV/$c^2$}}\xspace} 

\newcommand{\mevmass}{\ensuremath{\text{MeV/$c$}^2}\xspace}
\newcommand{\gevmass}{\ensuremath{\text{GeV/$c$}^2}\xspace}

\newcommand{\coupling}{\ensuremath{\text{GeV}^{-1}}\xspace}
\newcommand{\energy}{\ensuremath{\text{GeV}}\xspace}
\newcommand{\mass}{\ensuremath{\text{GeV/$c$}^2}\xspace}
\newcommand{\masssq}{\ensuremath{\text{GeV}^2\text{/$c$}^4}\xspace}
\newcommand{\epem}{\ensuremath{e^+e^-}\xspace}
\newcommand{\epemtoalp}{\ensuremath{\epem\to\gamma a, a \to\gamma\gamma}\xspace}
\newcommand{\gagg}{\ensuremath{g_{a\gamma\gamma}}\xspace}
\newcommand{\sig}{\ensuremath{\mathcal{S}}\xspace}

\begin{document}

\title{Search for Axionlike Particles Produced in \epem Collisions at \belletwo}

\date{\today}

\input{pub003}

\begin{abstract} 
We present a search for the direct production of a light pseudoscalar $a$ decaying into two photons with the \belletwo detector at the \superkekb collider.
We search for the process \hbox{$\epemtoalp$} in the mass range \hbox{$\maminnu \,< m_a < \mamax$} using data corresponding to an integrated luminosity of \intlumwitherror. 
Light pseudoscalars interacting predominantly with standard model gauge bosons (so-called axionlike particles or ALPs) are frequently postulated in extensions of the standard model.
We find no evidence for ALPs and set 95\% confidence level upper limits on the coupling strength \gagg of ALPs to photons at the level of \hbox{$10^{-3}\,\coupling$}.
The limits are the most restrictive to date for \hbox{$0.2\,<\,m_a\,<\,1\,\gevmass$}.
\end{abstract}

\maketitle

Axions and axionlike particles (ALPs) are predicted by many extensions of the standard model~(SM)~\cite{Jaeckel:2010ni}.
They occur, for example, in most solutions of the strong $CP$ problem~\cite{PhysRevLett.38.1440}.
ALPs share the quantum numbers of axions, but differ in that their masses and couplings are independent.
ALPs with sub-\mevmass masses are interesting in the context of astrophysics and cosmology and are cold dark matter~(DM) candidates, whereas ALPs with \hbox{$\mathcal{O}$(1 \gevmass)} masses generally relate to several topics in particle physics~\cite{Preskill:1982cy, Abbott:1982af, Dine:1982ah}. 
Most notably, heavy ALPs can connect the SM particles to yet undiscovered DM particles~\cite{Dolan:2017osp}. 
ALPs that predominantly couple to $\gamma\gamma$, $\gamma Z^0$, and $Z^0Z^0$ are experimentally much less constrained than those that couple to gluons or fermions.
The latter interactions typically lead to flavor-changing processes that can be probed in rare decays~\cite{Dolan2015}.
In this Letter we will consider the case that the ALP $a$ predominantly couples to photons, with coupling strength \gagg, and has negligible coupling strength $g_{a\gamma Z}$ to a photon and a $Z^0$ boson, so that \hbox{$\mathcal{B}(a\to\gamma\gamma)\approx100$\%}; we follow the notation for couplings introduced in Ref.~\cite{Dolan:2017osp}. 
In the \mevmass to \gevmass mass range, the current best limits for ALPs with photon couplings are derived from a variety of experiments. 
These limits come from $\epem\to\gamma+\rm{invisible}$ and beam-dump experiments for light ALPs~\cite{Dolan:2017osp, Dobrich2016, Banerjee:2020fue}, from $\epem\to\gamma\gamma$~\cite{JAECKEL2016482, Knapen:2016moh} and coherent Primakoff production off a nuclear target~\cite{PhysRevLett.123.071801} for intermediate-mass ALPs, and from peripheral heavy-ion collisions~\cite{Sirunyan:2018fhl} for heavy ALPs.

We search for \hbox{$\epemtoalp$} in the ALP mass range \hbox{$\maminnu \,< m_a < \mamax$} in the three-photon final state.
The signature in the center-of-mass (c.m.) system is a monoenergetic photon recoiling against the $a\to\gamma\gamma$ decay.
The energy of the recoil photon is 
\begin{displaymath}
E_{\rm{recoil }\gamma}^{\rm{c.m.}} = \frac{s-m_a^2}{2\sqrt{s}},
\end{displaymath} 
\noindent where $\sqrt{s}$ is the c.m.\ collision energy.
We search for an ALP signal as a narrow peak in the squared recoil-mass distribution $\mrecoilsq = s - 2 \sqrt{s} E_{\rm{recoil }\gamma}^{\rm{c.m.}}$, or as a narrow peak in the squared-invariant-mass distribution \minvsq, computed using the two-photon system, depending on which provides the better sensitivity.
We note that in the future a larger \belletwo dataset will be available to calibrate the photon covariance matrix, which in turn will allow the use of kinematic fitting of the three photons to the known beam four-momentum, thus improving the sensitivity.
In our search range, the width of the ALP is negligible with respect to the experimental resolution, and the ALP lifetime is negligible, thus it decays promptly.
The dominant SM background process is $\epem\to\gamma\gamma\gamma$.
The analysis selection, fit strategy, and limit-setting procedures are optimized and verified based on Monte Carlo simulation, i.e. without looking at data events, to avoid experimenter's bias.

We use a data set corresponding to an integrated luminosity of \intlumfullwitherror~\cite{Abudinen:2019osb} collected with the \belletwo detector at the asymmetric-energy \epem collider \superkekb~\cite{Akai:2018mbz}, which is located at the KEK laboratory in Tsukuba, Japan.
Data were collected at the c.m.\ energy of the $\Upsilon(4S)$ resonance ($\sqrt{s}=10.58$\,GeV) from April to July 2018.  
The energies of the electron and positron beams are 7\,\energy and 4\,\energy, respectively, resulting in a boost of $\beta\gamma= 0.28$ of the c.m.\ frame relative to the laboratory frame.
We use a randomly chosen subset of the data, approximately 10\%, to validate the selection, and we then discard it from the final data sample.
The remaining data set is used for the search and corresponds to an integrated luminosity of \intlumwitherror.

The \belletwo detector consists of several subdetectors arranged around the beam pipe in a cylindrical structure~\cite{Abe:2010gxa, Kou:2018nap}.
Only the components that are relevant to this analysis are described below.
Photons are measured and identified in the electromagnetic calorimeter (ECL) consisting of CsI(Tl) crystals.
The ECL provides both an energy and a timing measurement.
A superconducting solenoid situated outside of the calorimeter provides a 1.5\,T magnetic field.  
Charged-particle tracking is done using a silicon vertex detector (VXD) and a central drift chamber (CDC). 
Only one azimuthal octant of the VXD was present during the 2018 operations.
The $z$~axis of the laboratory frame coincides with that of the solenoid and its positive direction is approximately that of the incoming electron beam.
The polar angle $\theta$ is measured with respect to this direction.
Events are selected only by the hardware trigger, and no further software trigger selection is applied.
Trigger energy thresholds are very low and no vetoes for abundant QED scattering processes are applied.

We use \textsc{Babayaga@nlo}~\cite{CARLONICALAME2000459, CARLONICALAME200116, BALOSSINI2006227, BALOSSINI2008209} to generate SM background processes $\epem\to \epem(\gamma)$,  $\epem\to\gamma\gamma(\gamma)$.
We use \textsc{Phokhara9}~\cite{PhysRevD.97.016006} to generate SM background processes $\epem\to P\gamma(\gamma)$, where $P$ is a SM pseudoscalar meson $(\pi^0, \eta, \eta')$.
This includes production via the radiative decay of the intermediate vector resonances $\rho, \omega$, and $\phi$. 
The largest pseudoscalar background contribution for this analysis comes from $\epem\to\omega\gamma, \omega\to\pi^0\gamma$ with a boosted $\pi^0$ decaying into overlapping photons.
We use the same generators to calculate the cross sections of the respective processes.
We use \textsc{\hbox{MadGraph5}}~\cite{Alwall2014} to simulate signal events, including the effects of initial-state radiation (ISR) in event kinematics~\cite{Li:2018qnh}, for different hypotheses for $m_a$ in step sizes approximately equal to the signal resolution in our search range.

We use \textsc{Geant4}~\cite{AGOSTINELLI2003250} to simulate the interactions of particles in the detector, taking into account the nominal detector geometry and simulated beam-backgrounds adjusted to match the measured beam conditions.
We use the \belletwo software framework~\cite{Kuhr2018} to reconstruct and analyze events.

All selection criteria are chosen to maximize the Punzi figure of merit for $5\,\sigma$ discovery~\cite{Punzi:2003bu}.
Quantities are defined in the laboratory frame unless otherwise specified.
Photon candidates are reconstructed from ECL clusters with no associated charged tracks.
We select events with at least three photon candidates with energy $E_{\gamma}$ above 0.65\,\energy (for \hbox{$m_a > 4$\,\mass}) or 1.0\,\energy (for \hbox{$m_a \leq 4$\,\mass}).
This ALP-mass-dependent threshold is used to avoid shaping effects on the background distribution in the mass fit range.
The following selection variables are not dependent on the ALP mass.
All three photon candidates must be reconstructed with polar angles $37.3 < \theta_{\gamma} < 123.7\,^{\circ}$.
This polar-angle region provides the best calorimeter energy resolution, avoids regions close to detector gaps, and offers the lowest beam background levels.
If more than three photons pass the selection criteria, we select the three most energetic ones and the additional photons are ignored in the calculation of any variables. This occurs in fewer than 0.2\% of all events.
We reduce contamination from beam backgrounds by requiring that each photon detection time $t_i$ is compatible with the average weighted photon time

\begin{displaymath}
\bar{t}= \frac{\sum_{i=1}^{3} (t_i/\Delta t_i^2)}{\sum_{j=1}^{3}(1/\Delta t_j^2)},
\end{displaymath} 

\noindent where $\Delta t_i$ is the energy-dependent timing range that includes 99\% of all signal photons, and is between $3$~ns (high $E_{\gamma}$) and $15$~ns (low $E_{\gamma}$).
The requirement is $\vert (t_i-\bar{t})/\Delta t_i\vert<10$, which is insensitive to global time offsets.
The invariant mass $M_{\gamma\gamma\gamma}$ of the three-photon system must satisfy $0.88\sqrt{s} \leq M_{\gamma\gamma\gamma} \leq 1.03\sqrt{s}$ to eliminate kinematically unbalanced events coming from cosmic rays, beam-gas backgrounds, or two-photon production.
We reject events that have tracks originating from the interaction region to suppress background from $\epem\to\epem\gamma$.
We require a $\theta_{\gamma}$ separation between any two photons of $\Delta\theta_{\gamma} > 0.014\,\mathrm{rad}$, or an azimuthal angle separation of $\Delta\phi_{\gamma} > 0.400\,\mathrm{rad}$ to reduce background from photon conversions outside of the tracking detectors.
Following a data-sideband analysis using $M_{\gamma\gamma\gamma} < 0.88\,\sqrt{s}$, we additionally apply a loose selection, based on a multivariate shower-shape classifier that uses multiple Zernike moments~\cite{ZERNIKE1934689}, on the most isolated of the three photons.
This criterion reduces the number of clusters produced by neutral hadrons and by particles that do not originate from the interaction point.
The selection procedure results in three ALP candidates per event from all possible combinations of the three selected photons.

\begin{figure}[!htb]
\center{
\includegraphics[width=8.5cm]
{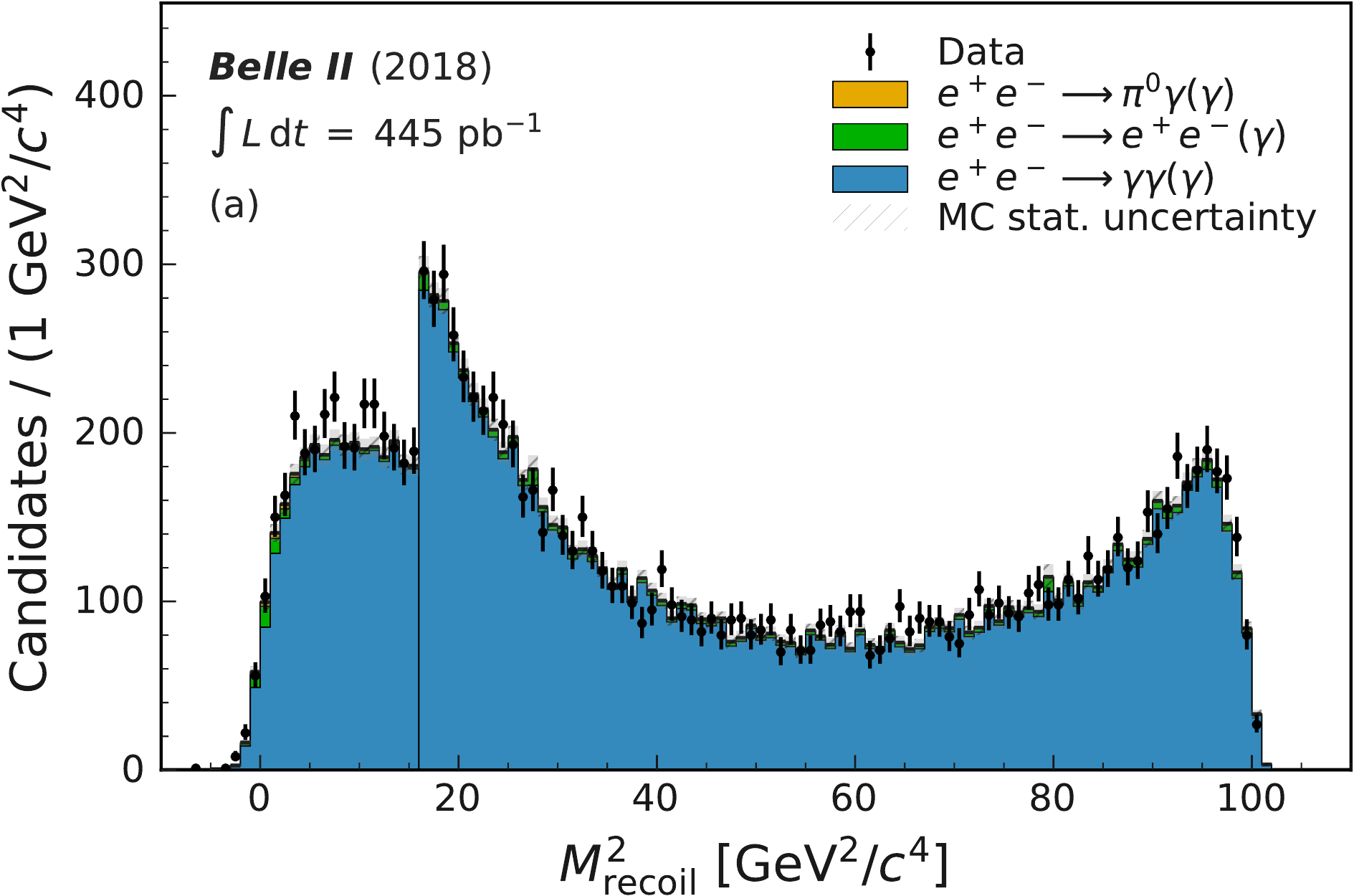}}
\center{\includegraphics[width=8.5cm]
{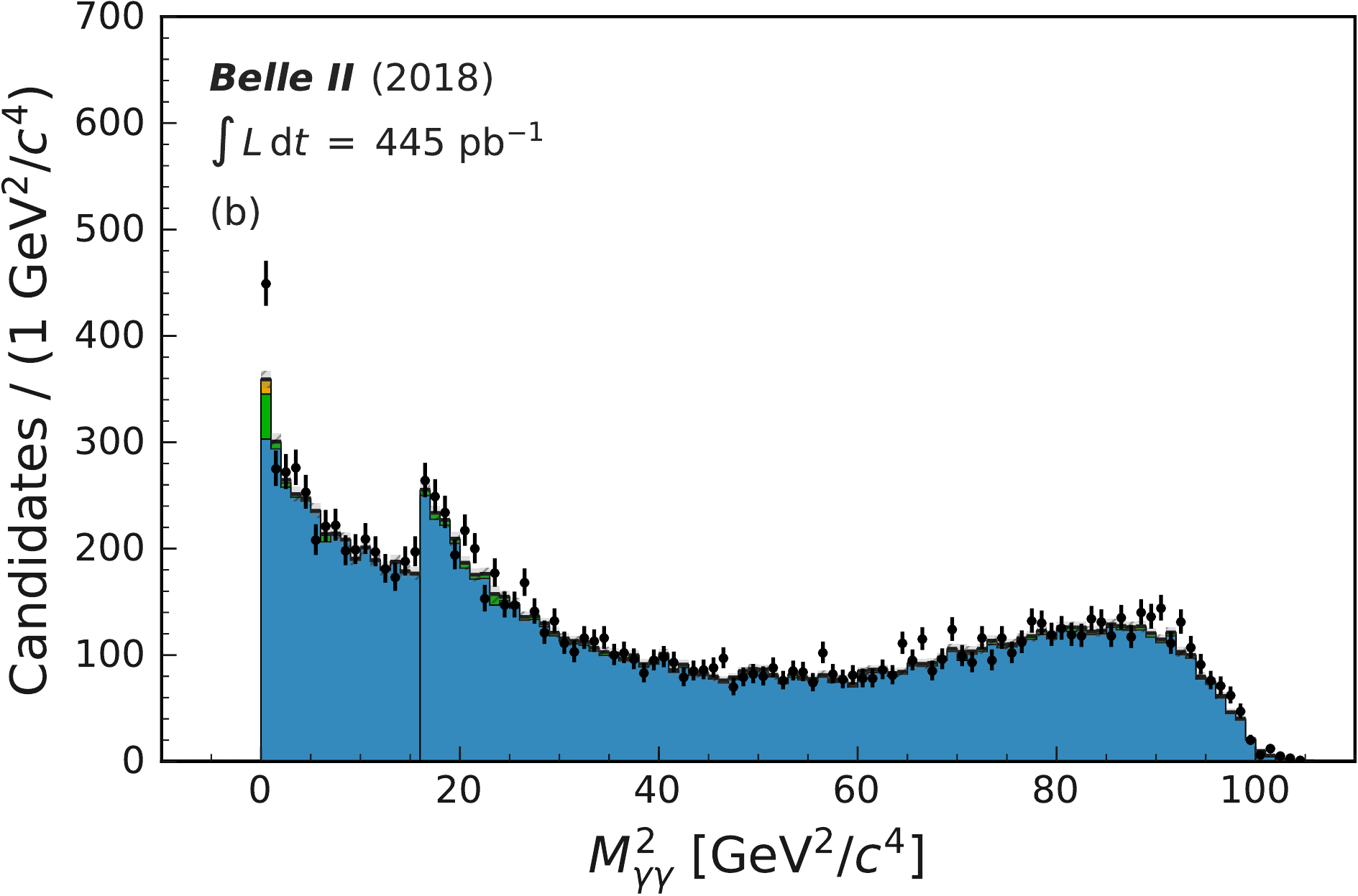}
\llap{\shortstack{%
        \includegraphics[width=4.75cm]{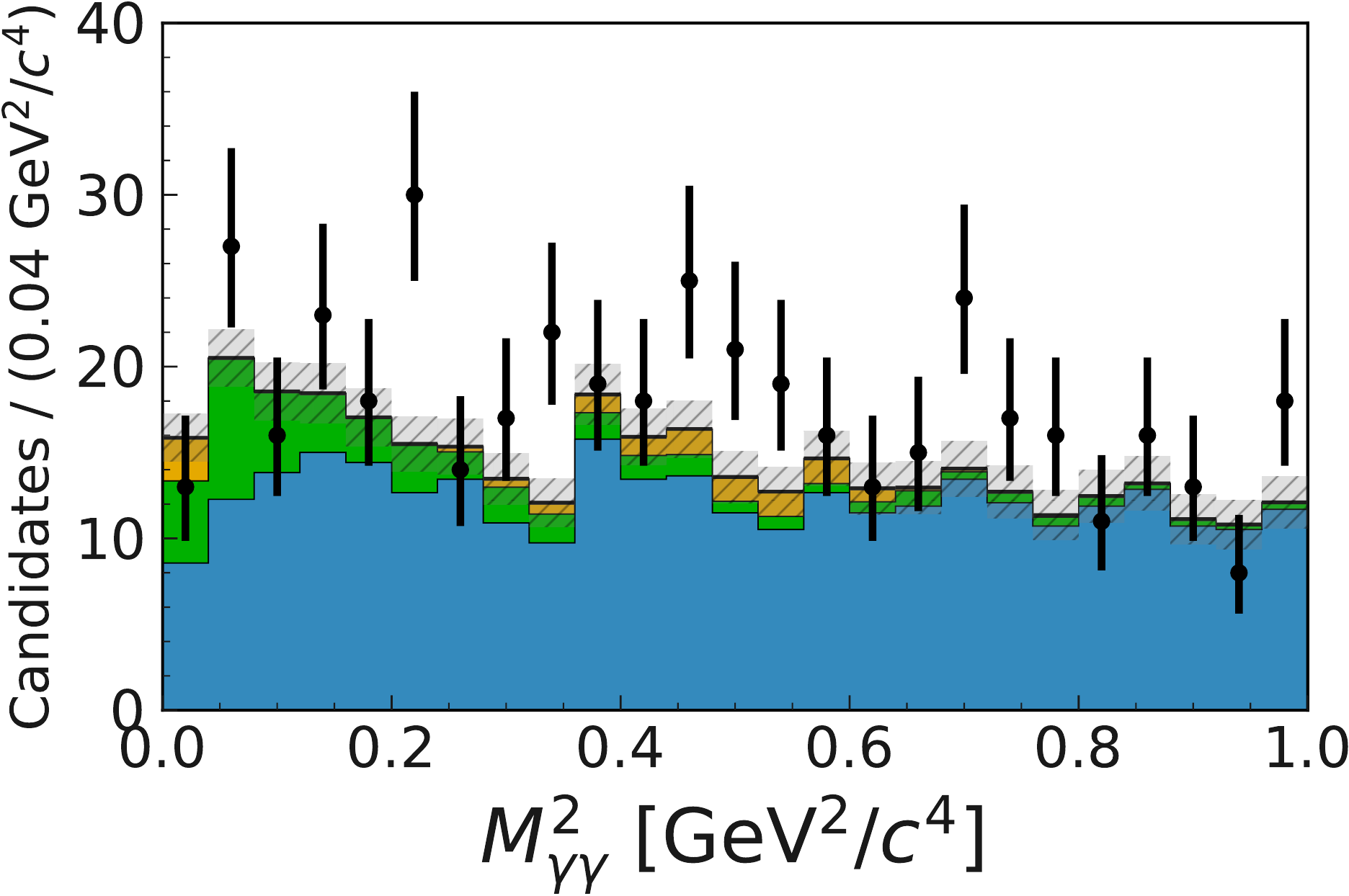}\\
        \rule{0ex}{2.0cm}%
      }
  \rule{0.1cm}{0ex}}}
\caption{\label{fig:distributions} \mrecoilsq distribution (a) and \minvsq distribution (b) together with the stacked contributions from the different simulated SM background samples. For $M^2 \leq 16$~$\masssq$, the selection is $E_{\gamma} > 1.0$~GeV; for $M^2 > 16$~$\masssq$, it is $E_{\gamma} > 0.65$~GeV. Simulation is normalized to luminosity. The inset in (b) shows an enlargement of the low-mass region \hbox{$M_{\gamma\gamma}^2 < 1\,\masssq$}.}
\end{figure}

\begin{figure}[!htb]
\center{\includegraphics[width=8.5cm]
{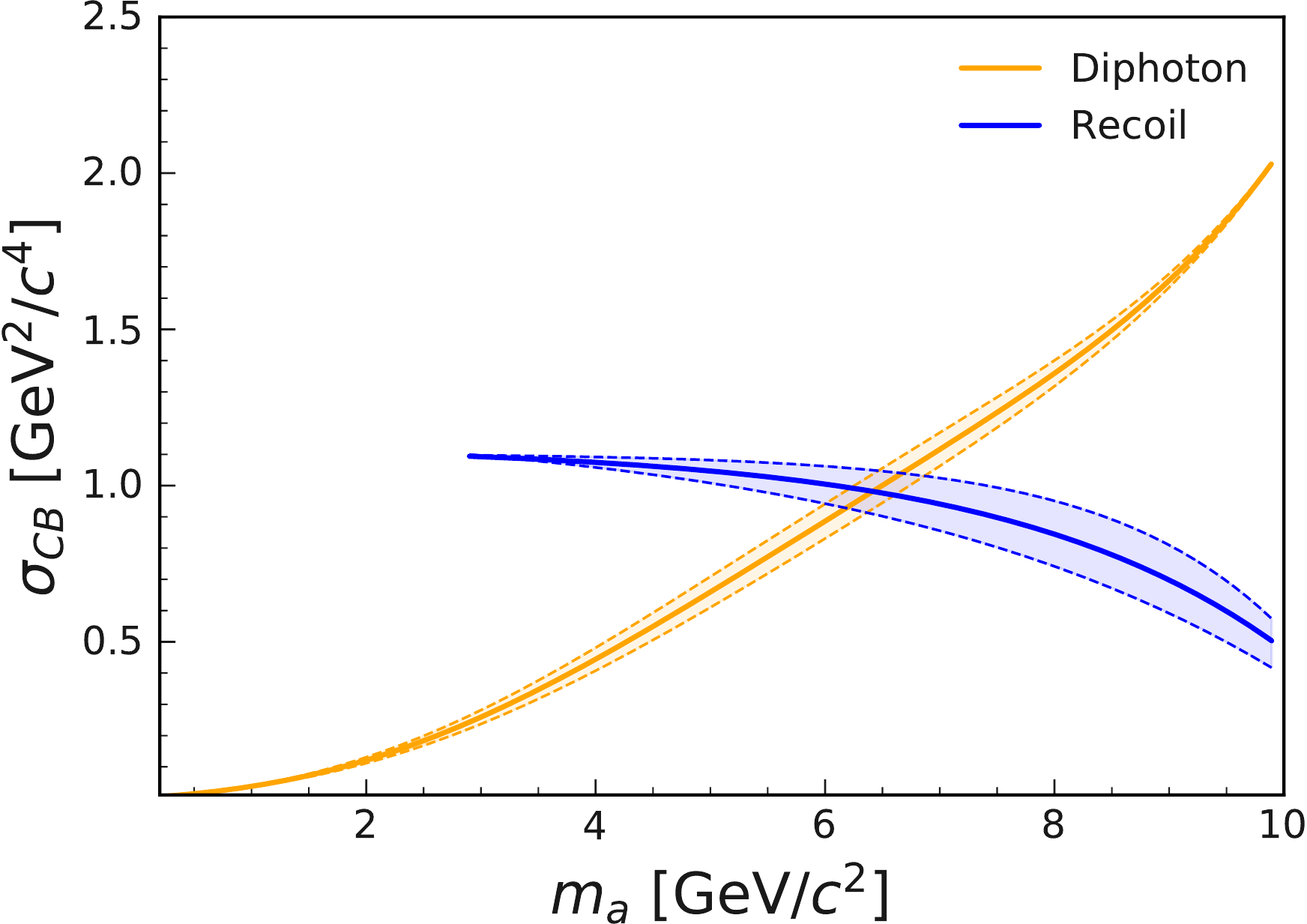}
\llap{\shortstack{%
        \includegraphics[width=3.9cm]{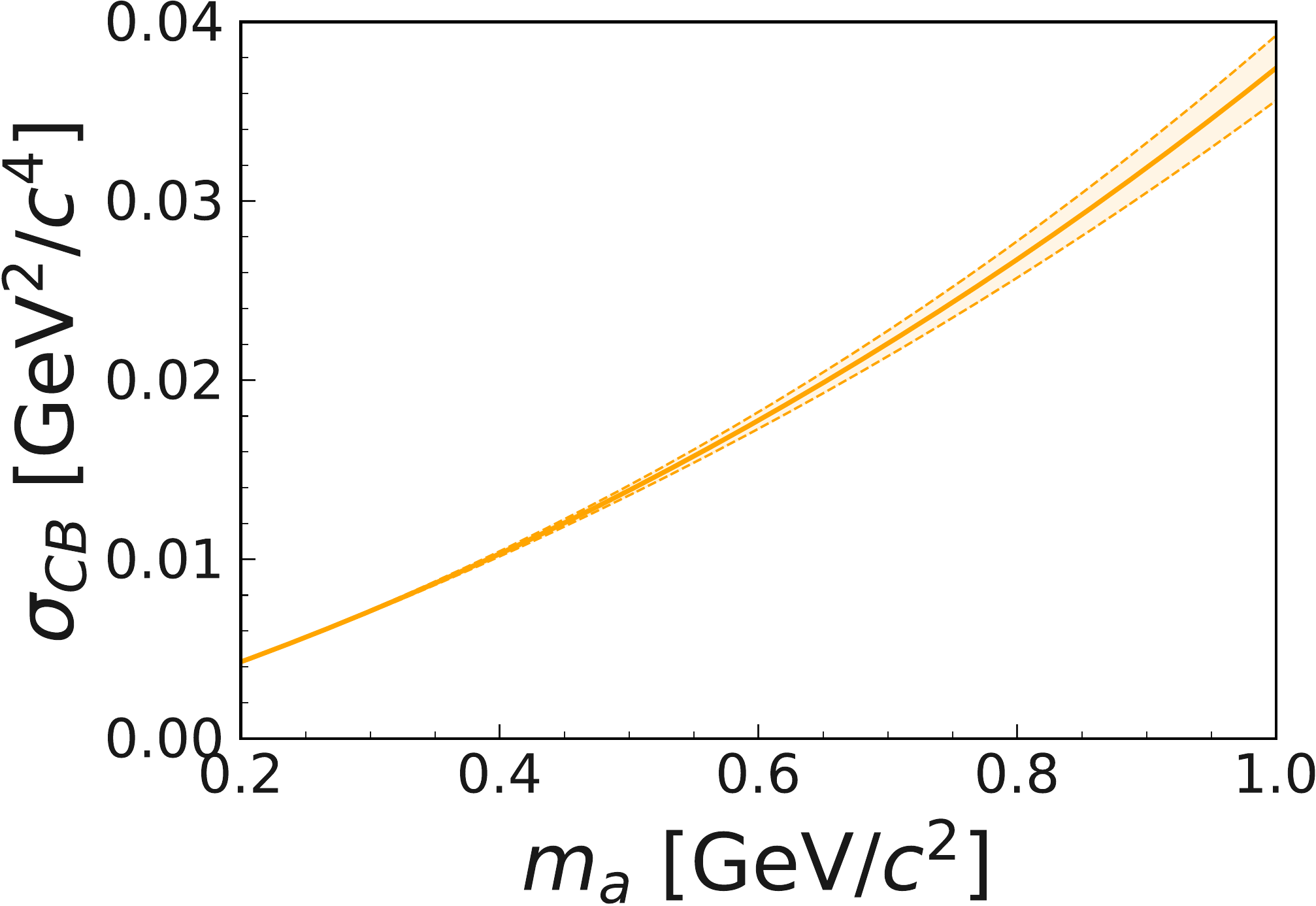}\\
        \rule{0ex}{3cm}%
      }
  \rule{3.25cm}{0ex}}}
\caption{\label{fig:resolution} \minvsq and \mrecoilsq resolutions with uncertainty as a function of ALP mass $m_a$.  The inset shows an enlargement of the low-mass region $m_a < 1\,\mass$.}
\end{figure}

The resulting \mrecoilsq and \minvsq distributions are shown in Fig.\,\ref{fig:distributions} together with the stacked contributions from the luminosity-normalized simulated samples of SM backgrounds.
The expected background distributions are dominated by $\epem\to\gamma\gamma\gamma$ with a small contribution from $\epem\to\epem\gamma$ due to tracking inefficiencies.
We find contributions from cosmic rays, assessed in data-taking periods without colliding beams, neither significant nor peaking in photon energy or invariant mass. 
The data shape agrees well with simulation except for a small and localized excess seen in the low-mass region $M_{\gamma\gamma}^2 < 1\,\masssq$. 
The excess is broad (see the inset in Fig.\,\ref{fig:distributions} (b)) and not consistent with an ALP signal, for which we expect a much smaller width in this region (see the inset in Fig.\,\ref{fig:resolution}).
As described later, the signal extraction does not directly depend on the background predictions because we fit the background only using data, thus any discrepancy between data and simulation has little impact on the result.
Triggers based on 1\,\energy threshold energy sums in the calorimeter barrel are found to have $\varepsilon_{\text{trg}}=1.0$ for the ALP selection, based upon studies of radiative Bhabha events.

The ALP selection efficiency is determined using large simulated signal samples, and varies smoothly between 20\% (low $m_a$) and 34\,\% (high $m_a$).
The number of candidates in data is $3.6\pm0.9$\% ($4.2\pm1.1$\%) higher than in the simulation for the $E_{\gamma}>0.65$\,\energy ($E_{\gamma}>1.0$\,\energy) selection.
No correction is applied and we assign the sum of the full difference and its uncertainty as a systematic uncertainty for the selection efficiency.
We assess the difference in the photon-energy reconstruction between data and simulation by using radiative muon-pair events in which we compare the predicted recoil energy calculated from the muon-pair momenta with the energy of the photon candidate.
We correct for the observed linear energy bias that ranges from 0 (low energy) to 0.5\% (high energy). 
We vary the energy selection by $\pm$1\% and the angular-separation selection by the approximate position resolution of $\pm$\,5\,mrad, and take the respective full difference in the signal selection efficiency with respect to the nominal selection as a systematic uncertainty.
We add these three uncertainties in quadrature assuming no correlations amongst them.
The total relative uncertainty due to the selection efficiency is approximately 5.5\% for ALP masses above 0.5\,\mass, and increases to approximately 8\% for the lightest ALP masses considered.
As additional systematic checks we vary the photon-timing selection by $\pm 1$ and the shower-shape classifier selection by $\pm\,5\%$ to account for possible between data and simulation samples, the invariant mass $M_{\gamma\gamma\gamma}$ selection by $\pm\,0.002$\,\mass to account for uncertainties in the beam energy, and the polar-angle-acceptance selection by propagating the effect of a $\pm2\,\mathrm{mm}$ shift of the interaction point relative to the calorimeter to account for maximal possible misalignment of the ECL.
For all of these checks, we find that they have a negligible effect on the signal selection efficiency, so we do not associate any systematic uncertainty with them.

We extract the signal yield as a function of $m_a$ by performing a series of independent binned maximum-likelihood fits.
We use 100 bins for each fit range.
The fits are performed in the range \hbox{$\fitrangeminnu < m_a < \fitrangetransition$} for the \minvsq spectrum, and in the range \hbox{$\fitrangetransitionnu < m_a <\fitrangemax$} for the \mrecoilsq spectrum.
The resolution of $M^2_{\gamma\gamma}$ worsens with increasing $m_a$, while that of \mrecoilsq improves with increasing $m_a$ (see Fig.\,\ref{fig:resolution}). 
The transition between $M_{\gamma\gamma}^2$ and \mrecoilsq fits is determined as the point of equal sensitivity obtained using background simulations.

The signal probability density function (PDF) has two components: a peaking contribution from correctly reconstructed signal photons and a combinatorial-background contribution from the other two combinations of photons.
We model the peaking contribution using a Crystal Ball (CB) function~\cite{Skwarnicki:1986xj}.
The mass-dependent CB parameters used in the fits to data are fixed to those obtained by fitting simulated events.
For the simulated \mrecoilsq distribution, the CB mean is found to be unbiased.
For the simulated \minvsq distribution, we observe a linear bias of the CB mean of about 0.5\% resulting from the combination of two photons with asymmetric reconstructed-energy distributions.
This bias is determined to have negligible impact on the signal yield and mass determination; therefore, no attempt to correct for it is made.
Combinatorial-background contributions from the wrong combinations of photons in signal events are taken into account by adding a mass-dependent, one-dimensional, smoothed kernel density estimation (KDE)~\cite{Cranmer:2000du} PDF obtained from signal simulation.
The fits are performed in steps of $m_a$ that correspond to half the CB width ($\sigma_{\rm{CB}}$) for the respective squared mass. This results in a total of \nfitsdiphoton fits to the \minvsq distribution and \nfitsrecoil fits to the \mrecoilsq distribution.
CB signal parameters are interpolated between the known simulated masses, and the KDE shape is taken from the simulation sample generated with the closest value of $m_a$ to that assumed in the fit.

The photon-energy resolution $\sigma(E_{\gamma})/E_{\gamma}$ in simulation is about 3\% for $E_{\gamma}=0.65$\,\energy and improves to about 2\% for $E_{\gamma}>1$\,\energy.
Using the same muon-pair sample as used for the photon-energy bias study, we find that the photon energy resolution in simulation is better than that in data by at most 30\% at low energies.
Therefore, we apply an energy-dependent additional resolution smearing to our simulated signal samples before determining the CB resolution parameter $\sigma_{\rm{CB}}$; we assume conservatively that the full observed difference between data and simulation is due to the photon-energy-resolution difference.
We assign half of the resulting mass-resolution difference as a systematic uncertainty.
The effect of a $\pm2\,\mathrm{mm}$ shift of the interaction point relative to the calorimeter is found to have a negligible impact on the mass resolution and is not included as a systematic uncertainty.

We describe the backgrounds by polynomials of the minimum complexity consistent with the data features.
Polynomials of second to fifth order are used: second for $0.2 < m_a \leq 0.5\,\gevmass$, fourth for $0.5 < m_a \leq 6.85\,\gevmass$, and fifth for $6.85 < m_a \leq 9.7\,\gevmass$.
The background polynomial parameters are not fixed by simulation but are free parameters of each data fit.
Each fit is performed in a mass range that corresponds to $-20$\,$\sigma_{\rm{CB}}$ to $+30$\,$\sigma_{\rm{CB}}$  for \minvsq, and $-25$\,$\sigma_{\rm{CB}}$  to $+25$\,$\sigma_{\rm{CB}}$  for \mrecoilsq.
In addition, the fit ranges are constrained between \hbox{$M^2_{\gamma\gamma} >0\,\masssq$} and \hbox{$M^2_{\text{recoil}} < 100.5\,\masssq$}.
The choice of the order of background polynomial and fit range is optimized based on the following conditions: giving a reduced $\chi^2$ close to one, providing locally smooth fit results, and being consistent with minimal variations between adjacent fit ranges.
Peaking backgrounds from $\epem\to P\gamma$ are very small compared to the expected statistical uncertainty on the signal yield and found to be modeled adequately by the polynomial background PDF.

\begin{figure}[!htb]
\center{\includegraphics[width=8.5cm]
{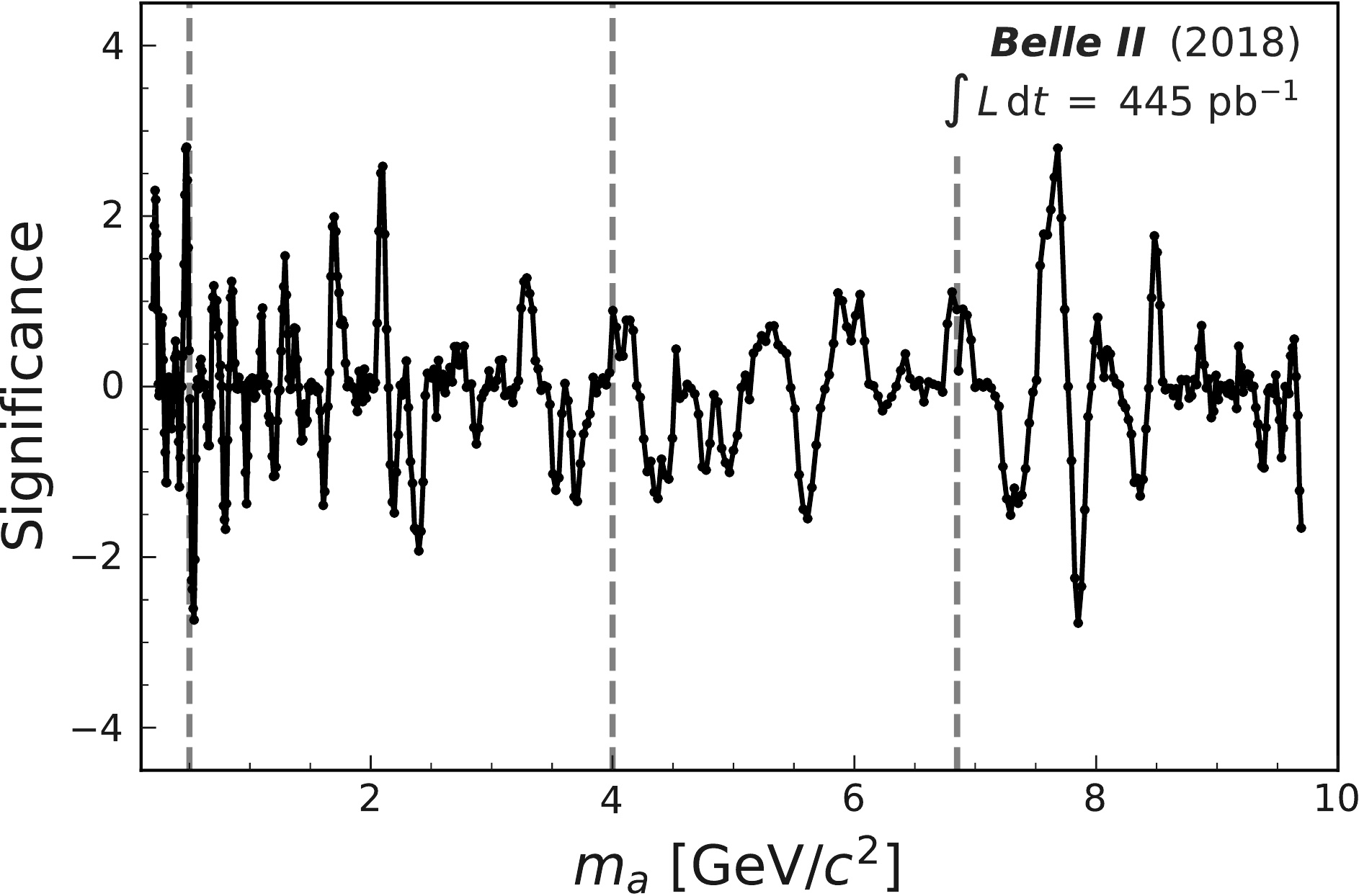}}
\caption{\label{fig:significance}  Local signal significance \sig multiplied by the sign of the signal yield, including systematic uncertainties, as a function of ALP mass $m_a$. The vertical dashed lines indicate (from left to right) changes in the default background PDF (0.5\,\mass), in the photon energy selection criteria (4.0\,\mass), and in the invariant-mass determination method (\fitrangetransition).}
\end{figure}

The systematic uncertainties due to the signal efficiency and the signal mass resolution are included as Gaussian nuisance parameters with a width equal to the systematic uncertainty.
The systematic uncertainty due to the background shape, which is the dominant source of systematic uncertainty, is estimated by repeating all fits with alternative fit ranges changed by $\pm 5\sigma_{\rm{CB}}$ and with the polynomial orders modified by $\pm 1$.
For each mass value $m_a$, we report the smallest of all signal significance values determined from each background model.
The local significance including systematic uncertainties is given by $\sig = \sqrt{2\ln(\mathcal{L}/\mathcal{L}_{\text{bkg}})}$, where $\mathcal{L}$ is the maximum likelihood for the fit, and $\mathcal{L}_{\text{bkg}}$ is the likelihood for a fit to the background-only hypothesis.
The local significances, multiplied by the sign of the signal yield, are shown in Fig.\,\ref{fig:significance}.
The largest local significance, including systematic uncertainties, is found near $m_a=0.477$\,\mass with a value of $\sig=2.8\,\sigma$.

\begin{figure}[!htb]
\center{\includegraphics[width=8.5cm]
{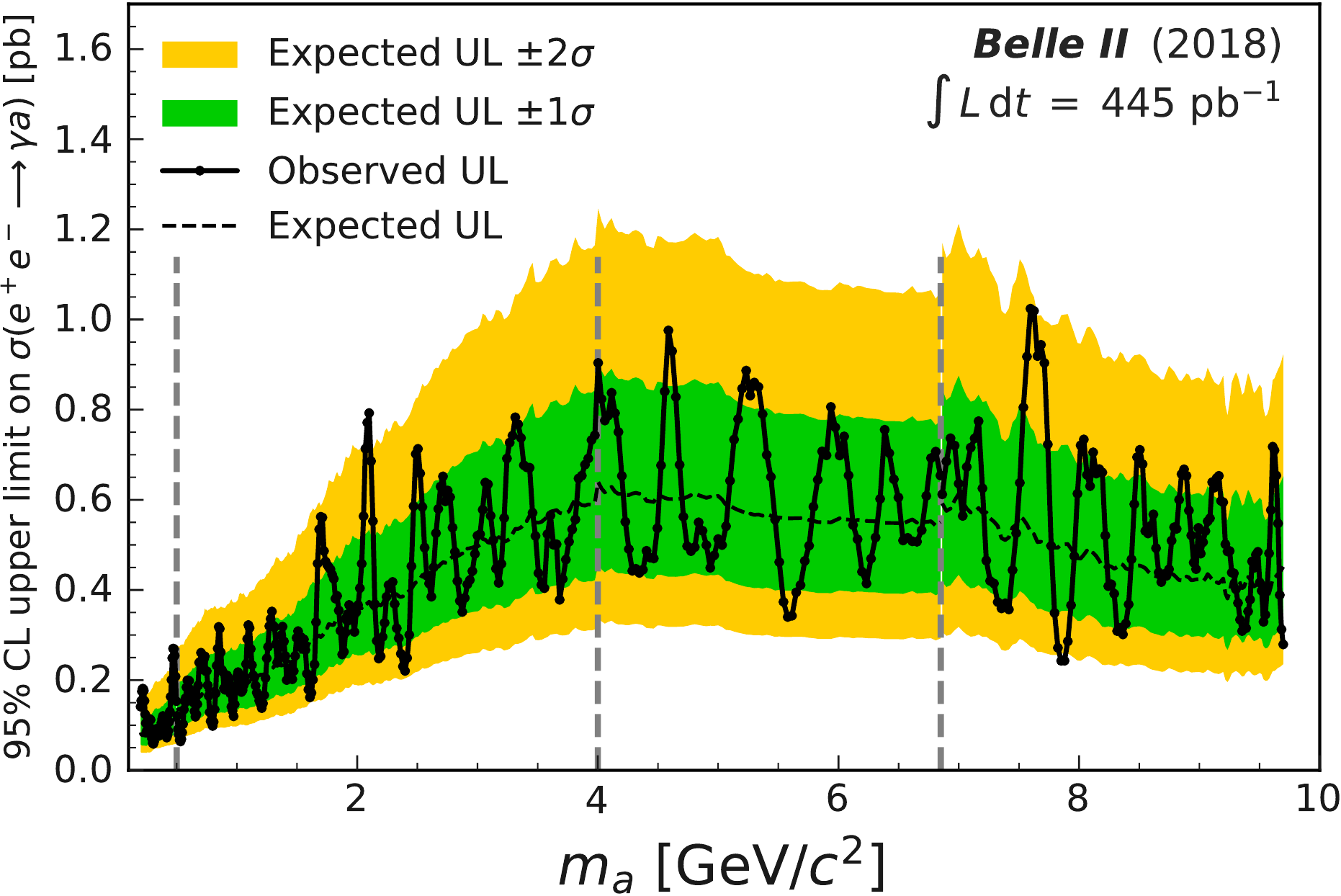}}
\caption{\label{fig:brazilband} Expected and observed upper limits (95\% C.L.) on the ALP cross section $\sigma_a$. The vertical dashed lines are the same as those in Fig.\,\ref{fig:significance}.}
\end{figure}

By dividing the signal yield by the signal efficiency and the integrated luminosity, we obtain the ALP cross section $\sigma_a$.
We compute the 95\% confidence level (C.L.) upper limits on $\sigma_a$ as a function of $m_a$  using a one-sided frequentist profile-likelihood method~\cite{Cowan:2010js}.
For each $m_a$ fit result, we report the least stringent of all 95\% C.L. upper limits determined from the variations of background model and fit range.
We convert the cross section limit to the coupling limit using
\begin{displaymath}
\sigma_a = \frac{\gagg^2\alpha_{\rm{QED}}}{24}\left(1-\frac{m_a^2}{s}\right)^3,
\end{displaymath} 
\noindent where $\alpha_{\rm{QED}}$ is the electromagnetic coupling~\cite{Dolan:2017osp}.
This calculation does not take into account any energy dependence of $\alpha_{\rm{QED}}$ and \gagg itself~\cite{Bauer:2017ris}.
An additional 0.2\% collision-energy uncertainty when converting $\sigma_a$ to \gagg results in a negligible additional systematic uncertainty.
Our median limit expected in the absence of a signal and the observed upper limits on $\sigma_a$ are shown in Fig.\,\ref{fig:brazilband}.
The observed upper limits on the photon couplings \gagg of ALPs, as well as existing constraints from previous experiments, are shown in Fig.\,\ref{fig:exclusion}.
Additional plots and numerical results can be found in the Supplemental Material~\cite{note:supmat}.
Our results provide the best limits for \hbox{$0.2 < m_a < 5\,\gevmass$}. 
This region of ALP parameter space is completely unconstrained by cosmological considerations~\cite{Depta:2020wmr}. 
The remaining mass region below \mamin is challenging to probe at colliders due to the poor spatial resolution of photons from highly boosted ALP decays, and irreducible peaking backgrounds from $\pi^0$ production.

\begin{figure}[!htb]
\center{\includegraphics[width=8.5cm]
{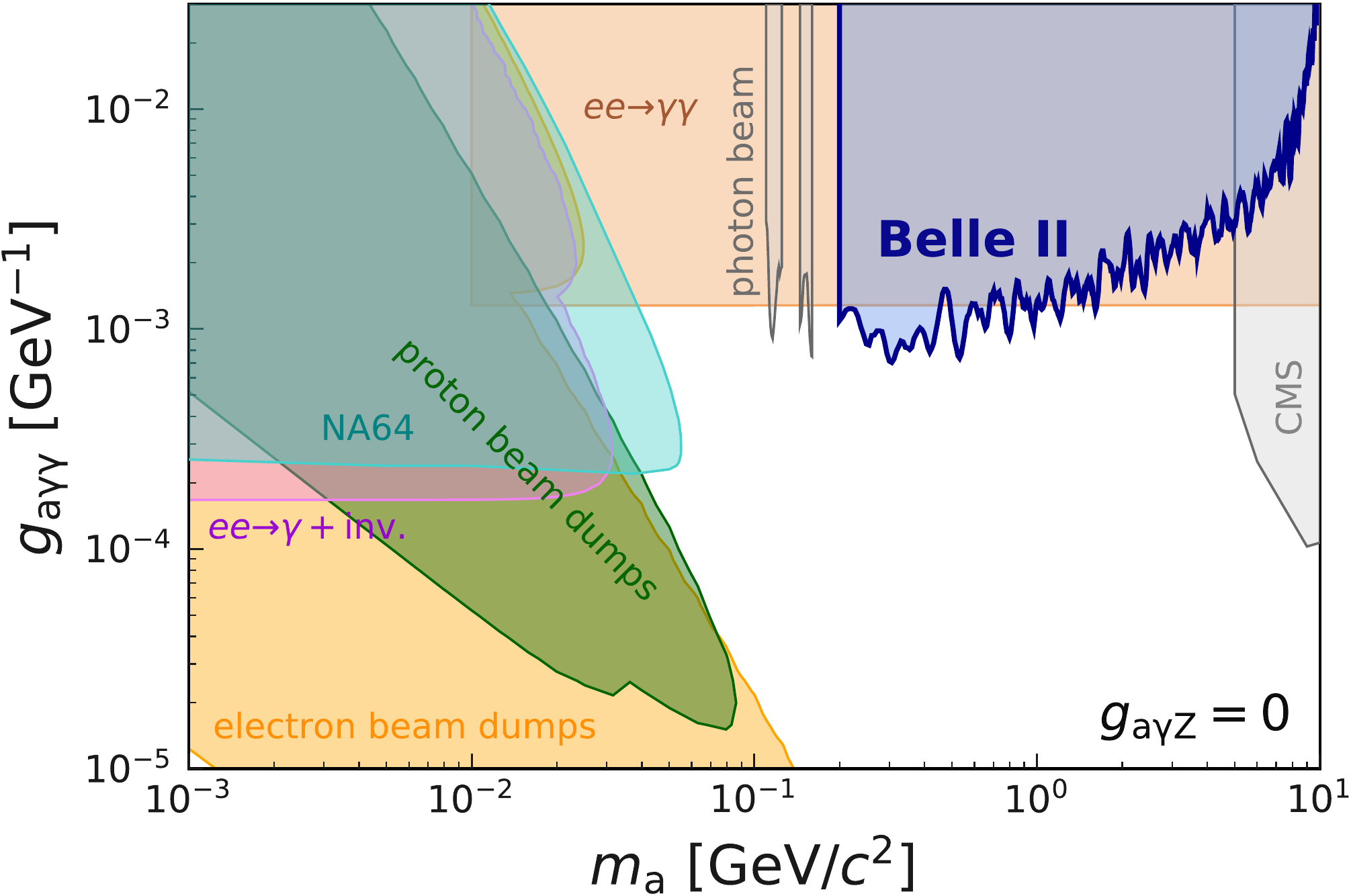}}
\caption{\label{fig:exclusion} Upper limit (95\% C.L.) on the ALP-photon coupling from this analysis and previous constraints from electron beam-dump experiments and $\epem\to\gamma+\rm{invisible}$~\cite{Dolan:2017osp, Banerjee:2020fue}, proton beam-dump experiments~\cite{Dobrich2016}, $\epem\to\gamma\gamma$~\cite{Knapen:2016moh}, a photon-beam experiment~\cite{PhysRevLett.123.071801}, and heavy-ion collisions~\cite{Sirunyan:2018fhl}.}
\end{figure}

In conclusion, we search for \hbox{$\epemtoalp$} in the ALP mass range \hbox{$\maminnu \,< m_a < \mamax$} using \belletwo data corresponding to an integrated luminosity of \intlum.
We do not observe any significant excess of events consistent with the signal process and set 95\%\,C.L. upper limits on the photon coupling \gagg at the level of \hbox{$10^{-3}\,\coupling$}. 
These limits, the first obtained for the fully reconstructed three-photon final state, are more restrictive than existing limits from LEP-II~\cite{Knapen:2016moh}.
In the future, with increased luminosity, \belletwo is expected to improve the sensitivity to \gagg by more than one order of magnitude~\cite{Dolan:2017osp}.

\input{acknowledgements.tex}



%


\end{document}


\title{Supplementary information:\\
Search for Axion-Like Particles produced in \epem collisions at \belletwo}

\maketitle

\begin{figure}[!htb]
\center{\includegraphics[width=12cm]
{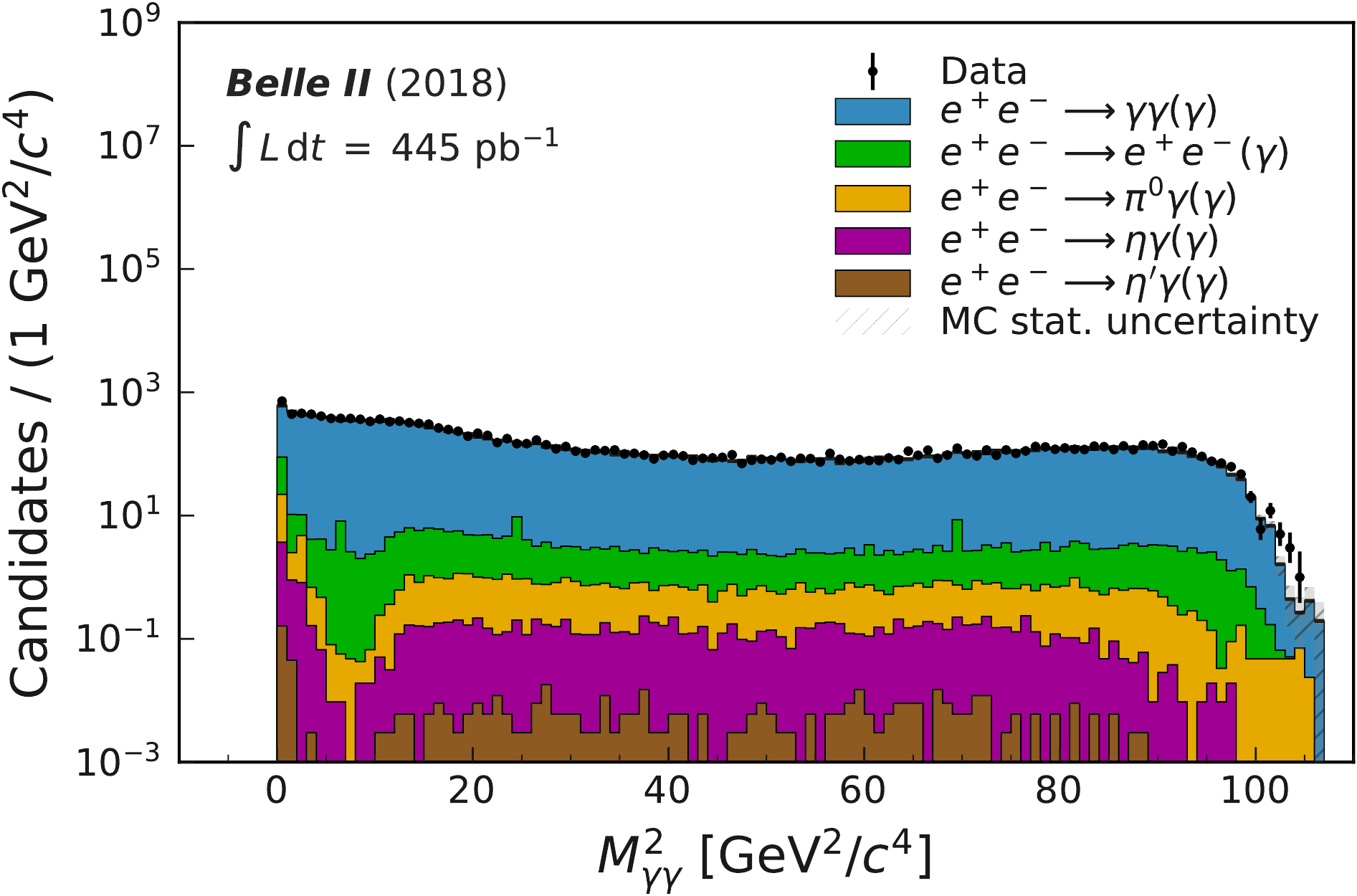}}
\caption{\label{fig:sup_mgg_650} Squared invariant mass distribution $M_{\gamma\gamma}$ together with the stacked contributions expected from the different simulated SM background samples for $E_{\gamma} > 0.65$\,GeV.}
\end{figure}

\begin{figure}[!htb]
\center{\includegraphics[width=12cm]
{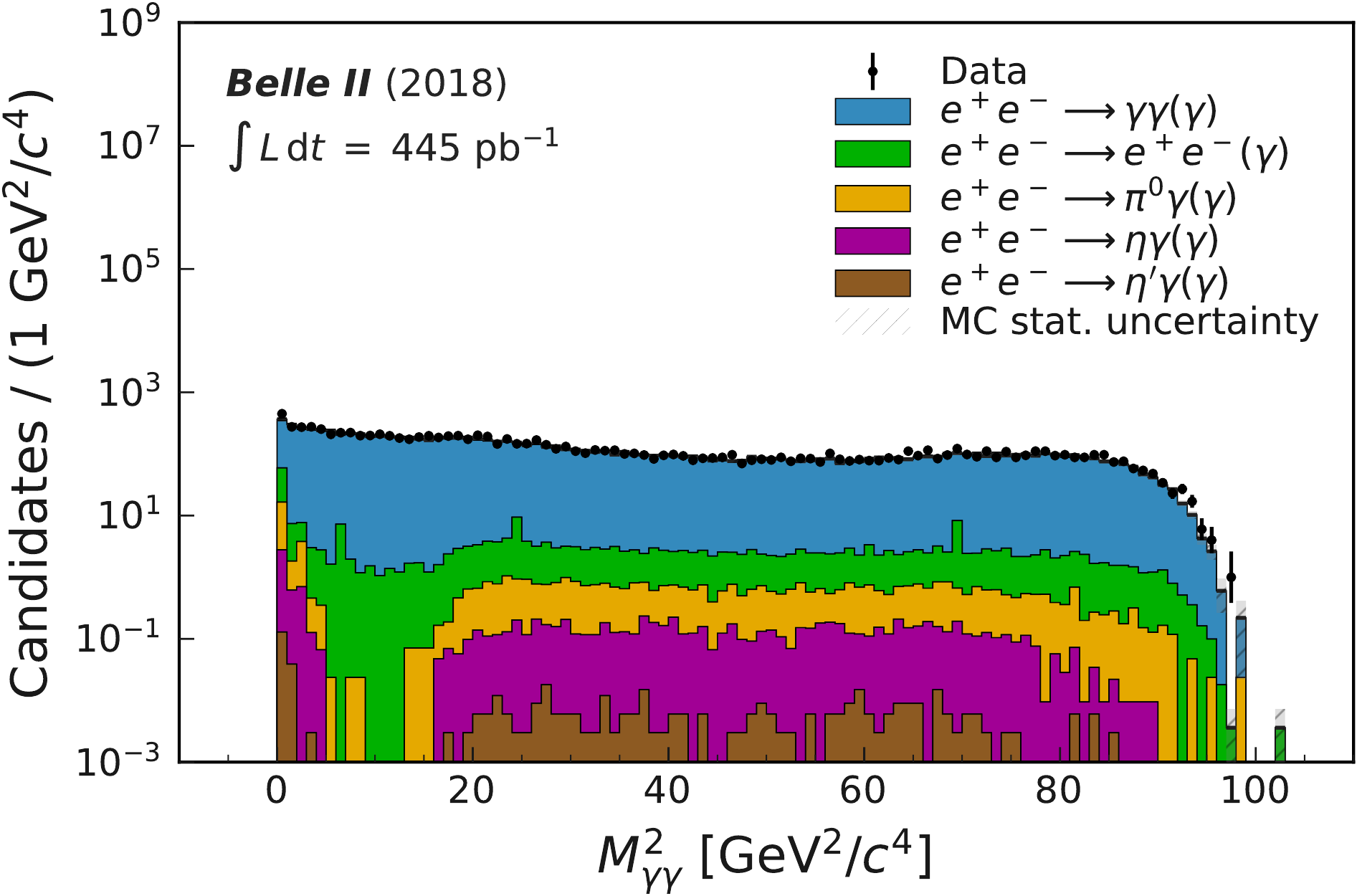}}
\caption{\label{fig:sup_mgg_1000} Squared invariant mass distribution $M_{\gamma\gamma}$ together with the stacked contributions expected from the different simulated SM background samples for $E_{\gamma} > 1.0$\,GeV.}
\end{figure}

\begin{figure}[!htb]
\center{\includegraphics[width=12cm]
{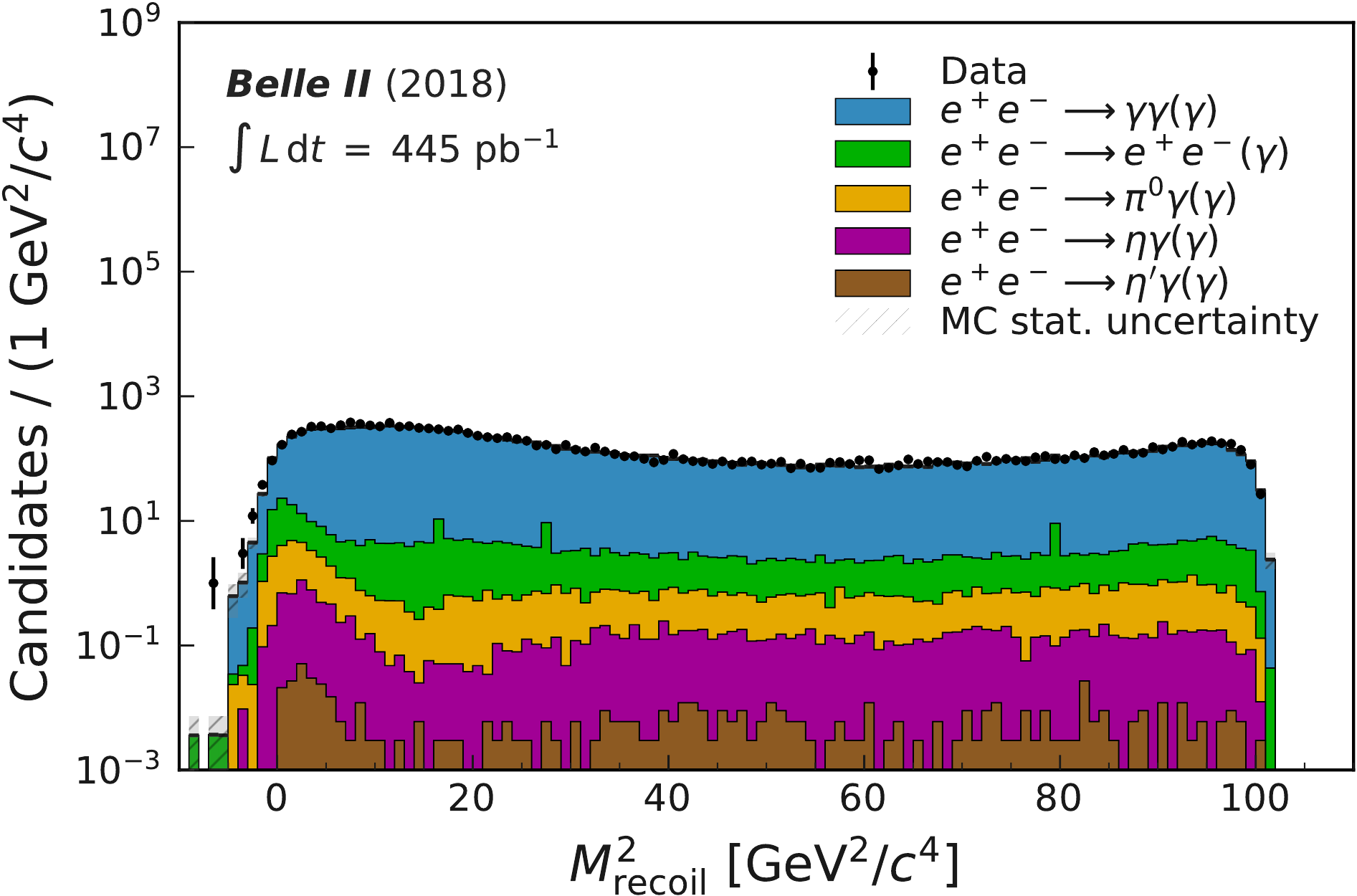}}
\caption{\label{fig:sup_mrecoil_650} Squared recoil-mass distribution $M_{\text{recoil}}^2$ together with the stacked contributions expected from the different simulated SM background samples for $E_{\gamma} > 0.65$\,GeV.}
\end{figure}

\begin{figure}[!htb]
\center{\includegraphics[width=12cm]
{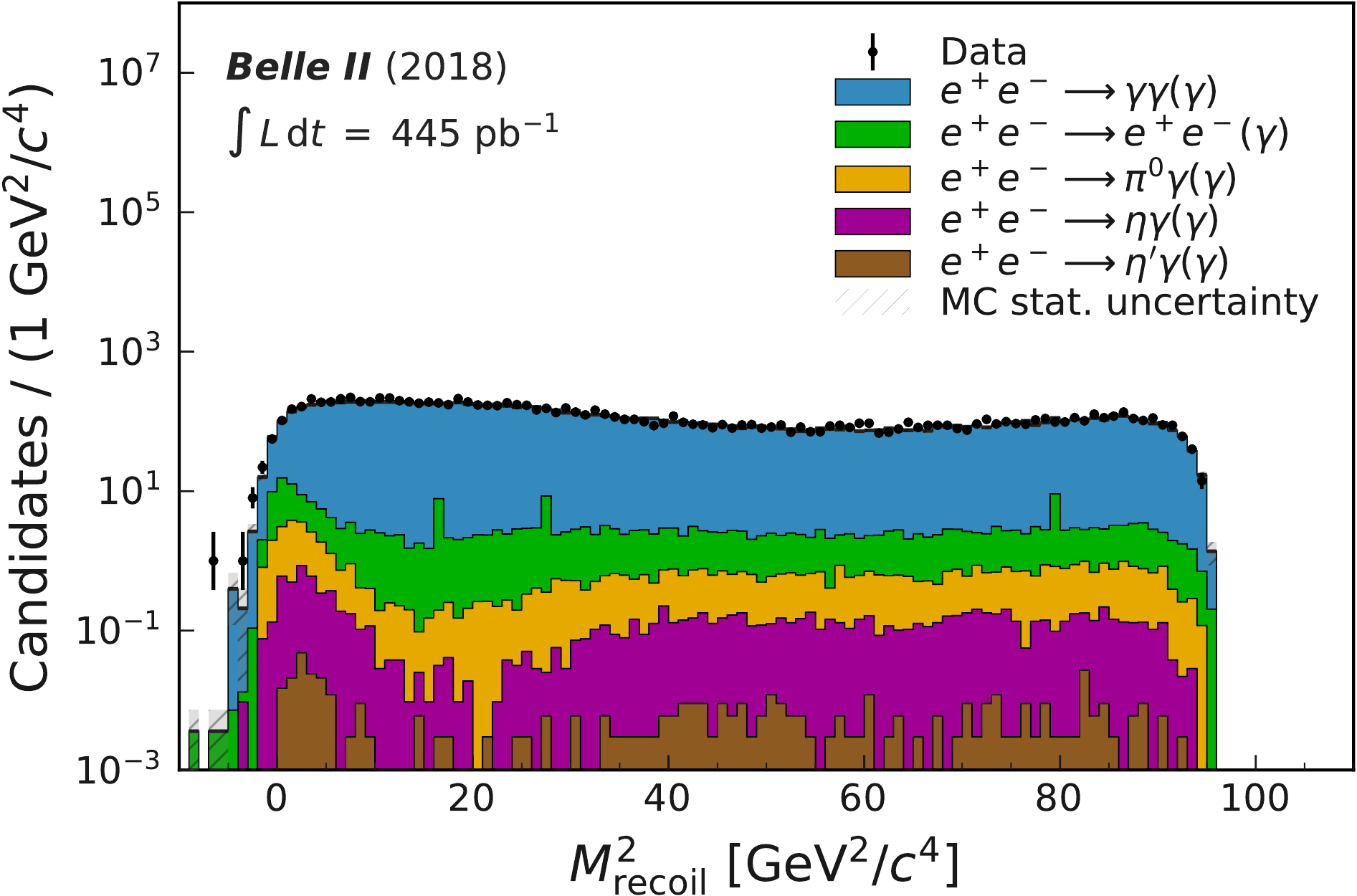}}
\caption{\label{fig:sup_mrecoil_1000} Squared recoil-mass distribution $M_{\text{recoil}}^2$ together with the stacked contributions expected from the different simulated SM background samples for $E_{\gamma} > 1.0$\,GeV.}
\end{figure}

\begin{figure}[!htb]
\center{\includegraphics[width=12cm]
{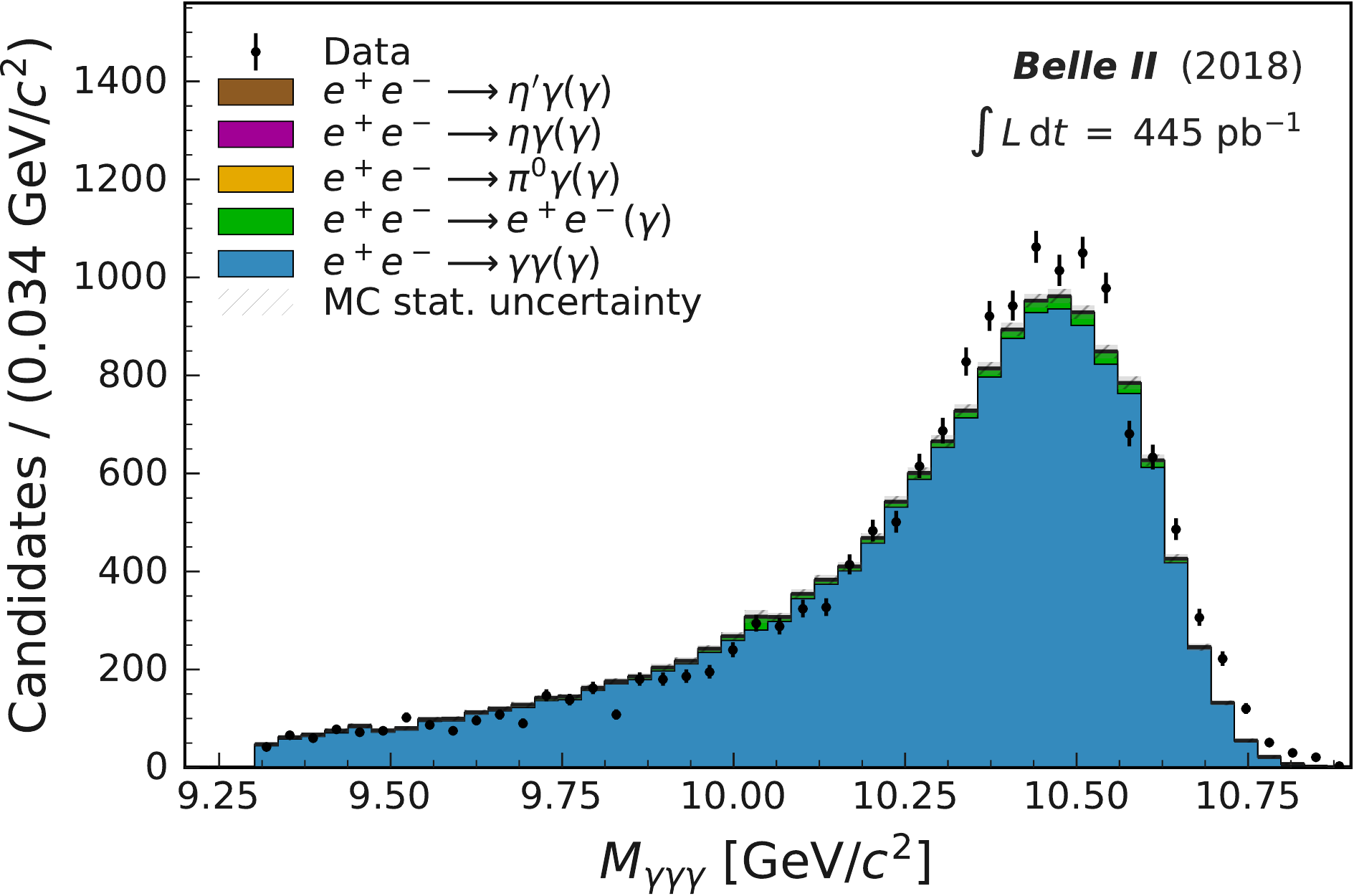}}
\caption{\label{fig:sup_mggg} Invariant mass distribution $M_{\gamma\gamma\gamma}$ together with the stacked contributions expected from the different simulated SM background samples for $E_{\gamma} > 0.65$\,GeV.}
\end{figure}

\begin{figure}[!htb]
\center{\includegraphics[width=12cm]
{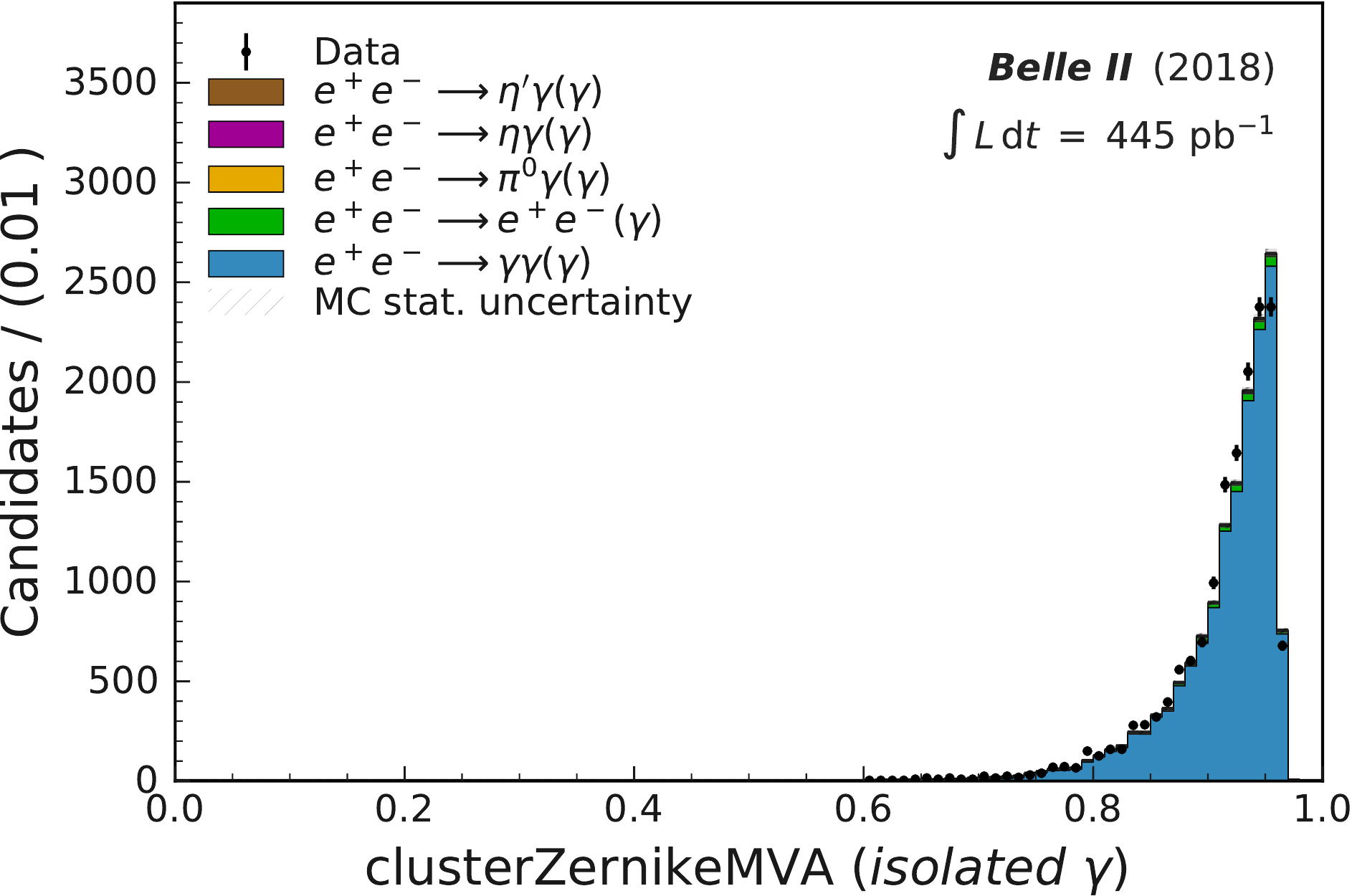}}
\caption{\label{fig:sup_zernikemva}  Zernike classifier distribution for the most isolated photon together with the stacked contributions expected from the different simulated SM background samples for $E_{\gamma} > 0.65$\,GeV.}
\end{figure}

\begin{figure}[!htb]
\center{\includegraphics[width=12cm]
{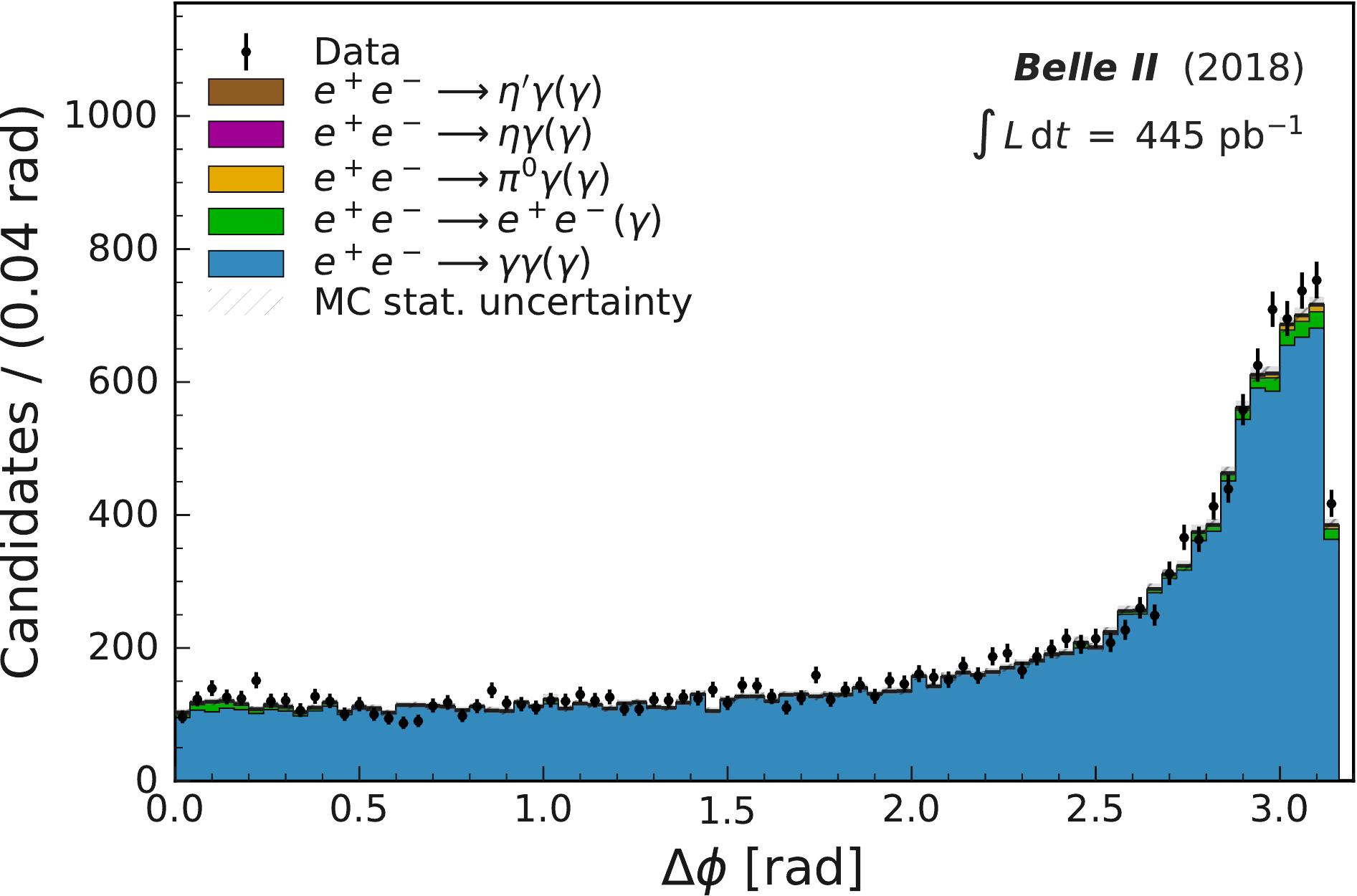}}
\caption{\label{fig:sup_deltaphi}  Azimuthal angle difference distribution for the most isolated photon together with the stacked contributions expected from the different simulated SM background samples for $E_{\gamma} > 0.65$\,GeV.}
\end{figure}

\begin{figure}[!htb]
\center{\includegraphics[width=12cm]
{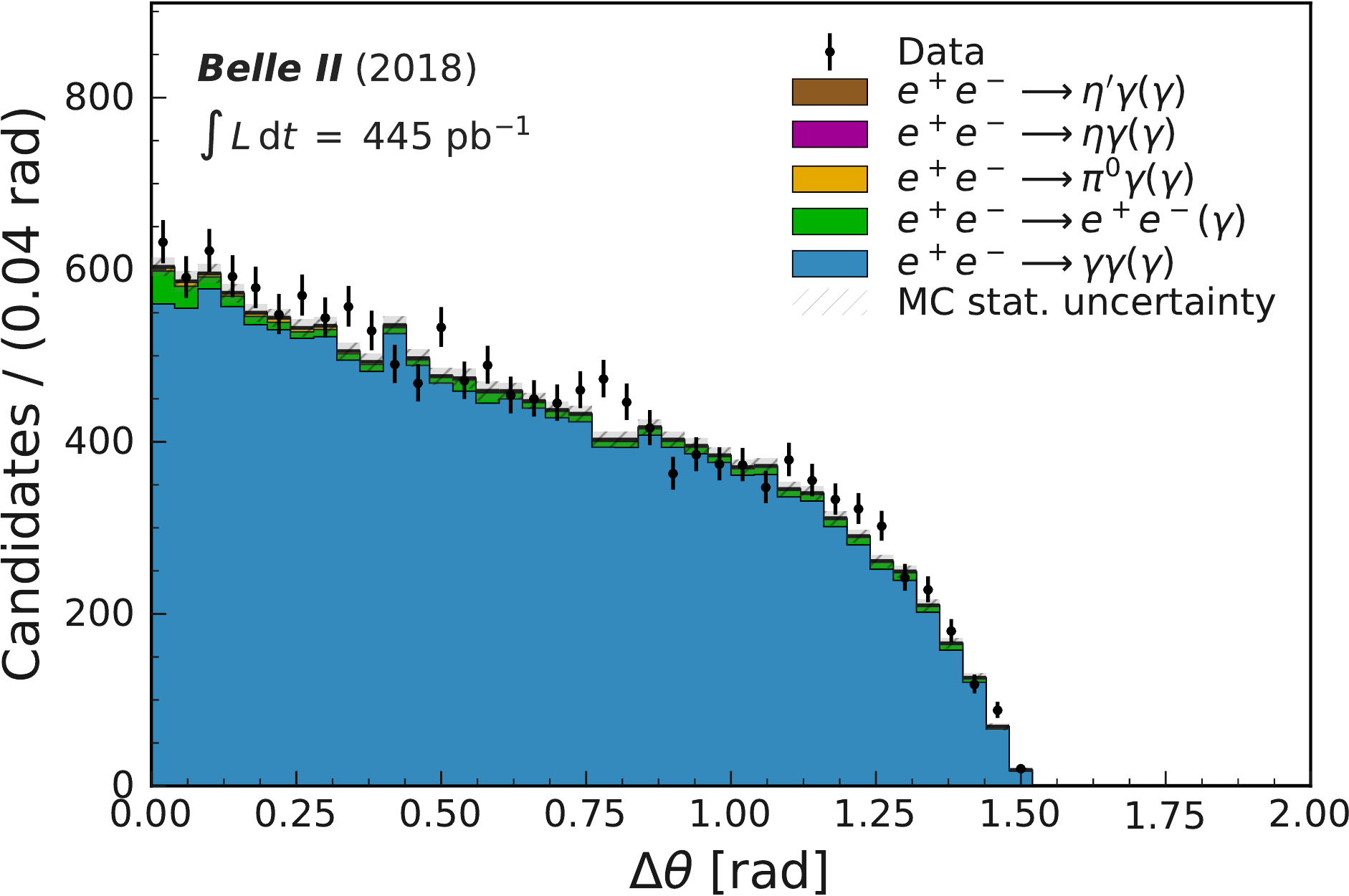}}
\caption{\label{fig:sup_deltatheta}  Polar angle difference distribution for the most isolated photon together with the stacked contributions expected from the different simulated SM background samples for $E_{\gamma} > 0.65$\,GeV.}
\end{figure}

\begin{figure}[!htb]
\center{\includegraphics[width=12cm]
{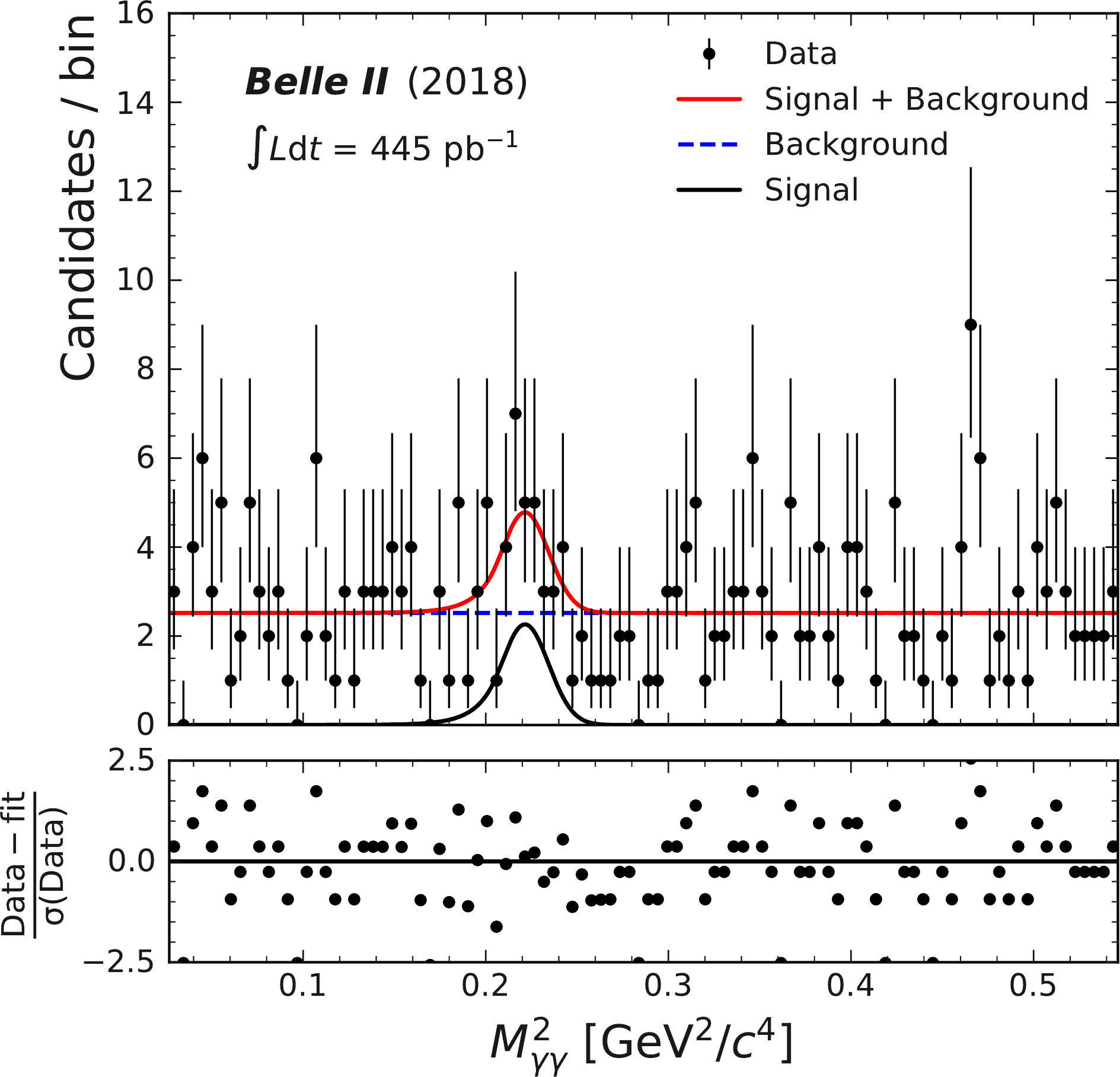}}
\caption{\label{fig:bestfit} Fit for $m_a=0.477$\,\mass with background PDF (constant, blue dashed), signal PDF (black) and total PDF (red). The signal peak corresponds to a local significance including systematic uncertainties of $\sig=2.8\,\sigma$.}
\end{figure}

Comparisons between data and simulated sample distributions of variables used in the selection and fit are given in Fig.~\ref{fig:sup_mgg_650}-\ref{fig:sup_deltatheta}. 
Fig.~\ref{fig:bestfit} shows the fit for $m_a=0.477$\,\mass that resulted in the largest significance observed.
We provide an additional textfile with numerical results the observed 95\% CL upper limit on the ALP cross section $\sigma_a$ (pb), and the observed 95\% CL upper limit on the photon coupling \gagg (GeV$^{-1}$) as a function of ALP mass $m_a$ (GeV).

%% file: pub003.tex
\newcommand{\instSinica}{Academia Sinica, Taipei 11529, Taiwan}
\newcommand{\instCPPM}{Aix Marseille Universit\'{e}, CNRS/IN2P3, CPPM, 13288 Marseille, France}
\newcommand{\instBeihang}{Beihang University, Beijing 100191, China}
\newcommand{\instBUAP}{Benemerita Universidad Autonoma de Puebla, Puebla 72570, Mexico}
\newcommand{\instBNL}{Brookhaven National Laboratory, Upton, New York 11973, U.S.A.}
\newcommand{\instBINP}{Budker Institute of Nuclear Physics SB RAS, Novosibirsk 630090, Russian Federation}
\newcommand{\instCMU}{Carnegie Mellon University, Pittsburgh, Pennsylvania 15213, U.S.A.}
\newcommand{\instCinvestavIPN}{Centro de Investigacion y de Estudios Avanzados del Instituto Politecnico Nacional, Mexico City 07360, Mexico}
\newcommand{\instPrague}{Faculty of Mathematics and Physics, Charles University, 121 16 Prague, Czech Republic}
\newcommand{\instChiangMai}{Chiang Mai University, Chiang Mai 50202, Thailand}
\newcommand{\instChiba}{Chiba University, Chiba 263-8522, Japan}
\newcommand{\instChonnam}{Chonnam National University, Gwangju 61186, South Korea}
\newcommand{\instConacyt}{Consejo Nacional de Ciencia y Tecnolog\'{\i}a, Mexico City 03940, Mexico}
\newcommand{\instDESY}{Deutsches Elektronen--Synchrotron, 22607 Hamburg, Germany}
\newcommand{\instDuke}{Duke University, Durham, North Carolina 27708, U.S.A.}
\newcommand{\instITAR}{Institute of Theoretical and Applied Research (ITAR), Duy Tan University, Hanoi 100000, Vietnam}
\newcommand{\instENEA}{ENEA Casaccia, I-00123 Roma, Italy}
\newcommand{\instEri}{Earthquake Research Institute, University of Tokyo, Tokyo 113-0032, Japan}
\newcommand{\instJuelich}{Forschungszentrum J\"{u}lich, 52425 J\"{u}lich, Germany}
\newcommand{\instFuJen}{Department of Physics, Fu Jen Catholic University, Taipei 24205, Taiwan}
\newcommand{\instFudan}{Key Laboratory of Nuclear Physics and Ion-beam Application (MOE) and Institute of Modern Physics, Fudan University, Shanghai 200443, China}
\newcommand{\instGoettingen}{II. Physikalisches Institut, Georg-August-Universit\"{a}t G\"{o}ttingen, 37073 G\"{o}ttingen, Germany}
\newcommand{\instGifu}{Gifu University, Gifu 501-1193, Japan}
\newcommand{\instSOKENDAI}{The Graduate University for Advanced Studies (SOKENDAI), Hayama 240-0193, Japan}
\newcommand{\instGyeongsang}{Gyeongsang National University, Jinju 52828, South Korea}
\newcommand{\instHanyang}{Department of Physics and Institute of Natural Sciences, Hanyang University, Seoul 04763, South Korea}
\newcommand{\instKEK}{High Energy Accelerator Research Organization (KEK), Tsukuba 305-0801, Japan}
\newcommand{\instJPARC}{J-PARC Branch, KEK Theory Center, High Energy Accelerator Research Organization (KEK), Tsukuba 305-0801, Japan}
\newcommand{\instHSE}{Higher School of Economics (HSE), Moscow 101000, Russian Federation}
\newcommand{\instIISER}{Indian Institute of Science Education and Research Mohali, SAS Nagar, 140306, India}
\newcommand{\instIITBhubaneswar}{Indian Institute of Technology Bhubaneswar, Satya Nagar 751007, India}
\newcommand{\instIITGuwahati}{Indian Institute of Technology Guwahati, Assam 781039, India}
\newcommand{\instIITHyderabad}{Indian Institute of Technology Hyderabad, Telangana 502285, India}
\newcommand{\instIITMadras}{Indian Institute of Technology Madras, Chennai 600036, India}
\newcommand{\instIndiana}{Indiana University, Bloomington, Indiana 47408, U.S.A.}
\newcommand{\instIHEPRussia}{Institute for High Energy Physics, Protvino 142281, Russian Federation}
\newcommand{\instHEPHYVienna}{Institute of High Energy Physics, Vienna 1050, Austria}
\newcommand{\instIHEPChina}{Institute of High Energy Physics, Chinese Academy of Sciences, Beijing 100049, China}
\newcommand{\instChennai}{Institute of Mathematical Sciences, Chennai 600113, India}
\newcommand{\instIPP}{Institute of Particle Physics (Canada), Victoria, British Columbia V8W 2Y2, Canada}
\newcommand{\instIOP}{Institute of Physics, Vietnam Academy of Science and Technology (VAST), Hanoi, Vietnam}
\newcommand{\instIFIC}{Instituto de Fisica Corpuscular, Paterna 46980, Spain}
\newcommand{\instFrascati}{INFN Laboratori Nazionali di Frascati, I-00044 Frascati, Italy}
\newcommand{\instNapoliINFN}{INFN Sezione di Napoli, I-80126 Napoli, Italy}
\newcommand{\instPadovaINFN}{INFN Sezione di Padova, I-35131 Padova, Italy}
\newcommand{\instPerugiaINFN}{INFN Sezione di Perugia, I-06123 Perugia, Italy}
\newcommand{\instPisaINFN}{INFN Sezione di Pisa, I-56127 Pisa, Italy}
\newcommand{\instRomaINFN}{INFN Sezione di Roma, I-00185 Roma, Italy}
\newcommand{\instRomaTreINFN}{INFN Sezione di Roma Tre, I-00146 Roma, Italy}
\newcommand{\instTorinoINFN}{INFN Sezione di Torino, I-10125 Torino, Italy}
\newcommand{\instTriesteINFN}{INFN Sezione di Trieste, I-34127 Trieste, Italy}
\newcommand{\instJAEA}{Advanced Science Research Center, Japan Atomic Energy Agency, Naka 319-1195, Japan}
\newcommand{\instMainz}{Johannes Gutenberg-Universit\"{a}t Mainz, Institut f\"{u}r Kernphysik, D-55099 Mainz, Germany}
\newcommand{\instGiessen}{Justus-Liebig-Universit\"{a}t Gie\ss{}en, 35392 Gie\ss{}en, Germany}
\newcommand{\instKarlsruhe}{Institut f\"{u}r Experimentelle Teilchenphysik, Karlsruher Institut f\"{u}r Technologie, 76131 Karlsruhe, Germany}
\newcommand{\instKennesaw}{Kennesaw State University, Kennesaw, Georgia 30144, U.S.A.}
\newcommand{\instKitasato}{Kitasato University, Sagamihara 252-0373, Japan}
\newcommand{\instKISTI}{Korea Institute of Science and Technology Information, Daejeon 34141, South Korea}
\newcommand{\instKorea}{Korea University, Seoul 02841, South Korea}
\newcommand{\instKSU}{Kyoto Sangyo University, Kyoto 603-8555, Japan}
\newcommand{\instKyotoU}{Kyoto University, Kyoto 606-8501, Japan}
\newcommand{\instKyungpook}{Kyungpook National University, Daegu 41566, South Korea}
\newcommand{\instLPI}{P.N. Lebedev Physical Institute of the Russian Academy of Sciences, Moscow 119991, Russian Federation}
\newcommand{\instLNNU}{Liaoning Normal University, Dalian 116029, China}
\newcommand{\instLMU}{Ludwig Maximilians University, 80539 Munich, Germany}
\newcommand{\instLuther}{Luther College, Decorah, Iowa 52101, U.S.A.}
\newcommand{\instMNITJaipur}{Malaviya National Institute of Technology Jaipur, Jaipur 302017, India}
\newcommand{\instMPP}{Max-Planck-Institut f\"{u}r Physik, 80805 M\"{u}nchen, Germany}
\newcommand{\instMPGHLL}{Semiconductor Laboratory of the Max Planck Society, 81739 M\"{u}nchen, Germany}
\newcommand{\instMcGill}{McGill University, Montr\'{e}al, Qu\'{e}bec, H3A 2T8, Canada}
\newcommand{\instMETU}{Middle East Technical University, 06531 Ankara, Turkey}
\newcommand{\instMEPhI}{Moscow Physical Engineering Institute, Moscow 115409, Russian Federation}
\newcommand{\instNagoya}{Graduate School of Science, Nagoya University, Nagoya 464-8602, Japan}
\newcommand{\instNagoyaKMI}{Kobayashi-Maskawa Institute, Nagoya University, Nagoya 464-8602, Japan}
\newcommand{\instNaraWu}{Nara Women's University, Nara 630-8506, Japan}
\newcommand{\instUNAM}{National Autonomous University of Mexico, Mexico City, Mexico}
\newcommand{\instNTUTaiwan}{Department of Physics, National Taiwan University, Taipei 10617, Taiwan}
\newcommand{\instNUUTaiwan}{National United University, Miao Li 36003, Taiwan}
\newcommand{\instKrakow}{H. Niewodniczanski Institute of Nuclear Physics, Krakow 31-342, Poland}
\newcommand{\instNiigata}{Niigata University, Niigata 950-2181, Japan}
\newcommand{\instNSU}{Novosibirsk State University, Novosibirsk 630090, Russian Federation}
\newcommand{\instOkinawa}{Okinawa Institute of Science and Technology, Okinawa 904-0495, Japan}
\newcommand{\instOsakaCity}{Osaka City University, Osaka 558-8585, Japan}
\newcommand{\instRCNP}{Research Center for Nuclear Physics, Osaka University, Osaka 567-0047, Japan}
\newcommand{\instPNNL}{Pacific Northwest National Laboratory, Richland, Washington 99352, U.S.A.}
\newcommand{\instPanjab}{Panjab University, Chandigarh 160014, India}
\newcommand{\instPeking}{Peking University, Beijing 100871, China}
\newcommand{\instPanjabPAU}{Punjab Agricultural University, Ludhiana 141004, India}
\newcommand{\instRIKEN}{Theoretical Research Division, Nishina Center, RIKEN, Saitama 351-0198, Japan}
\newcommand{\instXavier}{St. Francis Xavier University, Antigonish, Nova Scotia, B2G 2W5, Canada}
\newcommand{\instSeoul}{Seoul National University, Seoul 08826, South Korea}
\newcommand{\instShandong}{Shandong University, Jinan 250100, China}
\newcommand{\instSPU}{Showa Pharmaceutical University, Tokyo 194-8543, Japan}
\newcommand{\instSoochow}{Soochow University, Suzhou 215006, China}
\newcommand{\instSoongsil}{Soongsil University, Seoul 06978, South Korea}
\newcommand{\instLjubljanaJSI}{J. Stefan Institute, 1000 Ljubljana, Slovenia}
\newcommand{\instKyiv}{Taras Shevchenko National Univ. of Kiev, Kiev, Ukraine}
\newcommand{\instTata}{Tata Institute of Fundamental Research, Mumbai 400005, India}
\newcommand{\instTUM}{Department of Physics, Technische Universit\"{a}t M\"{u}nchen, 85748 Garching, Germany}
\newcommand{\instECUTUM}{Excellence Cluster Universe, Technische Universit\"{a}t M\"{u}nchen, 85748 Garching, Germany}
\newcommand{\instTelAviv}{Tel Aviv University, School of Physics and Astronomy, Tel Aviv, 69978, Israel}
\newcommand{\instToho}{Toho University, Funabashi 274-8510, Japan}
\newcommand{\instTohoku}{Department of Physics, Tohoku University, Sendai 980-8578, Japan}
\newcommand{\instTitech}{Tokyo Institute of Technology, Tokyo 152-8550, Japan}
\newcommand{\instTokyoMetropolitan}{Tokyo Metropolitan University, Tokyo 192-0397, Japan}
\newcommand{\instUAS}{Universidad Autonoma de Sinaloa, Sinaloa 80000, Mexico}
\newcommand{\instNapoliUNIVA}{Dipartimento di Agraria, Universit\`{a} di Napoli Federico II, I-80055 Portici (NA), Italy}
\newcommand{\instNapoliUNIV}{Dipartimento di Scienze Fisiche, Universit\`{a} di Napoli Federico II, I-80126 Napoli, Italy}
\newcommand{\instPadovaUNIV}{Dipartimento di Fisica e Astronomia, Universit\`{a} di Padova, I-35131 Padova, Italy}
\newcommand{\instPerugiaUNIV}{Dipartimento di Fisica, Universit\`{a} di Perugia, I-06123 Perugia, Italy}
\newcommand{\instPisaUNIV}{Dipartimento di Fisica, Universit\`{a} di Pisa, I-56127 Pisa, Italy}
\newcommand{\instRomaUNIV}{Universit\`{a} di Roma ``La Sapienza,'' I-00185 Roma, Italy}
\newcommand{\instRomaTreUNIV}{Dipartimento di Matematica e Fisica, Universit\`{a} di Roma Tre, I-00146 Roma, Italy}
\newcommand{\instTorinoUNIV}{Dipartimento di Fisica, Universit\`{a} di Torino, I-10125 Torino, Italy}
\newcommand{\instTriesteUNIV}{Dipartimento di Fisica, Universit\`{a} di Trieste, I-34127 Trieste, Italy}
\newcommand{\instMontreal}{Universit\'{e} de Montr\'{e}al, Physique des Particules, Montr\'{e}al, Qu\'{e}bec, H3C 3J7, Canada}
\newcommand{\instIJCLab}{Universit\'{e} Paris-Saclay, CNRS/IN2P3, IJCLab, 91405 Orsay, France}
\newcommand{\instIPHC}{Universit\'{e} de Strasbourg, CNRS, IPHC, UMR 7178, 67037 Strasbourg, France}
\newcommand{\instAdelaide}{Department of Physics, University of Adelaide, Adelaide, South Australia 5005, Australia}
\newcommand{\instBonn}{University of Bonn, 53115 Bonn, Germany}
\newcommand{\instUBC}{University of British Columbia, Vancouver, British Columbia, V6T 1Z1, Canada}
\newcommand{\instCincinnati}{University of Cincinnati, Cincinnati, Ohio 45221, U.S.A.}
\newcommand{\instFlorida}{University of Florida, Gainesville, Florida 32611, U.S.A.}
\newcommand{\instHamburg}{University of Hamburg, 20148 Hamburg, Germany}
\newcommand{\instHawaii}{University of Hawaii, Honolulu, Hawaii 96822, U.S.A.}
\newcommand{\instHeidelberg}{University of Heidelberg, 68131 Mannheim, Germany}
\newcommand{\instLjubljanaUniLJ}{Faculty of Mathematics and Physics, University of Ljubljana, 1000 Ljubljana, Slovenia}
\newcommand{\instLouisville}{University of Louisville, Louisville, Kentucky 40292, U.S.A.}
\newcommand{\instMalaya}{National Centre for Particle Physics, University Malaya, 50603 Kuala Lumpur, Malaysia}
\newcommand{\instLjubljanaUM}{University of Maribor, 2000 Maribor, Slovenia}
\newcommand{\instMelbourne}{School of Physics, University of Melbourne, Victoria 3010, Australia}
\newcommand{\instMississippi}{University of Mississippi, University, Mississippi 38677, U.S.A.}
\newcommand{\instUOM}{University of Miyazaki, Miyazaki 889-2192, Japan}
\newcommand{\instNovaGorica}{University of Nova Gorica, 5000 Nova Gorica, Slovenia}
\newcommand{\instPittsburgh}{University of Pittsburgh, Pittsburgh, Pennsylvania 15260, U.S.A.}
\newcommand{\instUSTC}{University of Science and Technology of China, Hefei 230026, China}
\newcommand{\instSAlabama}{University of South Alabama, Mobile, Alabama 36688, U.S.A.}
\newcommand{\instSCarolina}{University of South Carolina, Columbia, South Carolina 29208, U.S.A.}
\newcommand{\instSydney}{School of Physics, University of Sydney, New South Wales 2006, Australia}
\newcommand{\instTabuk}{Department of Physics, Faculty of Science, University of Tabuk, Tabuk 71451, Saudi Arabia}
\newcommand{\instUTokyo}{Department of Physics, University of Tokyo, Tokyo 113-0033, Japan}
\newcommand{\instIPMU}{Kavli Institute for the Physics and Mathematics of the Universe (WPI), University of Tokyo, Kashiwa 277-8583, Japan}
\newcommand{\instVictoria}{University of Victoria, Victoria, British Columbia, V8W 3P6, Canada}
\newcommand{\instVPI}{Virginia Polytechnic Institute and State University, Blacksburg, Virginia 24061, U.S.A.}
\newcommand{\instWayneState}{Wayne State University, Detroit, Michigan 48202, U.S.A.}
\newcommand{\instYamagata}{Yamagata University, Yamagata 990-8560, Japan}
\newcommand{\instYerevan}{Alikhanyan National Science Laboratory, Yerevan 0036, Armenia}
\newcommand{\instYonsei}{Yonsei University, Seoul 03722, South Korea}
\affiliation{\instCPPM}
\affiliation{\instBNL}
\affiliation{\instBINP}
\affiliation{\instCinvestavIPN}
\affiliation{\instPrague}
\affiliation{\instChiangMai}
\affiliation{\instChiba}
\affiliation{\instConacyt}
\affiliation{\instDESY}
\affiliation{\instDuke}
\affiliation{\instITAR}
\affiliation{\instEri}
\affiliation{\instJuelich}
\affiliation{\instFuJen}
\affiliation{\instFudan}
\affiliation{\instGoettingen}
\affiliation{\instGifu}
\affiliation{\instSOKENDAI}
\affiliation{\instGyeongsang}
\affiliation{\instHanyang}
\affiliation{\instKEK}
\affiliation{\instJPARC}
\affiliation{\instHSE}
\affiliation{\instIISER}
\affiliation{\instIITBhubaneswar}
\affiliation{\instIITHyderabad}
\affiliation{\instIITMadras}
\affiliation{\instIndiana}
\affiliation{\instHEPHYVienna}
\affiliation{\instIHEPChina}
\affiliation{\instIPP}
\affiliation{\instIOP}
\affiliation{\instIFIC}
\affiliation{\instFrascati}
\affiliation{\instNapoliINFN}
\affiliation{\instPadovaINFN}
\affiliation{\instPerugiaINFN}
\affiliation{\instPisaINFN}
\affiliation{\instRomaINFN}
\affiliation{\instRomaTreINFN}
\affiliation{\instTorinoINFN}
\affiliation{\instTriesteINFN}
\affiliation{\instJAEA}
\affiliation{\instMainz}
\affiliation{\instGiessen}
\affiliation{\instKarlsruhe}
\affiliation{\instKISTI}
\affiliation{\instKorea}
\affiliation{\instKyungpook}
\affiliation{\instLPI}
\affiliation{\instLNNU}
\affiliation{\instLMU}
\affiliation{\instLuther}
\affiliation{\instMNITJaipur}
\affiliation{\instMPP}
\affiliation{\instMcGill}
\affiliation{\instMEPhI}
\affiliation{\instNagoya}
\affiliation{\instNagoyaKMI}
\affiliation{\instNaraWu}
\affiliation{\instNTUTaiwan}
\affiliation{\instNUUTaiwan}
\affiliation{\instNiigata}
\affiliation{\instNSU}
\affiliation{\instOkinawa}
\affiliation{\instRCNP}
\affiliation{\instPNNL}
\affiliation{\instPanjab}
\affiliation{\instPeking}
\affiliation{\instPanjabPAU}
\affiliation{\instRIKEN}
\affiliation{\instSeoul}
\affiliation{\instSPU}
\affiliation{\instSoochow}
\affiliation{\instSoongsil}
\affiliation{\instLjubljanaJSI}
\affiliation{\instKyiv}
\affiliation{\instTata}
\affiliation{\instTUM}
\affiliation{\instTelAviv}
\affiliation{\instToho}
\affiliation{\instTohoku}
\affiliation{\instTitech}
\affiliation{\instTokyoMetropolitan}
\affiliation{\instUAS}
\affiliation{\instNapoliUNIVA}
\affiliation{\instNapoliUNIV}
\affiliation{\instPadovaUNIV}
\affiliation{\instPerugiaUNIV}
\affiliation{\instPisaUNIV}
\affiliation{\instRomaUNIV}
\affiliation{\instRomaTreUNIV}
\affiliation{\instTorinoUNIV}
\affiliation{\instTriesteUNIV}
\affiliation{\instMontreal}
\affiliation{\instIJCLab}
\affiliation{\instIPHC}
\affiliation{\instAdelaide}
\affiliation{\instBonn}
\affiliation{\instUBC}
\affiliation{\instCincinnati}
\affiliation{\instHawaii}
\affiliation{\instHeidelberg}
\affiliation{\instLjubljanaUniLJ}
\affiliation{\instLouisville}
\affiliation{\instMalaya}
\affiliation{\instLjubljanaUM}
\affiliation{\instMelbourne}
\affiliation{\instMississippi}
\affiliation{\instPittsburgh}
\affiliation{\instUSTC}
\affiliation{\instSAlabama}
\affiliation{\instSCarolina}
\affiliation{\instSydney}
\affiliation{\instUTokyo}
\affiliation{\instIPMU}
\affiliation{\instVictoria}
\affiliation{\instVPI}
\affiliation{\instWayneState}
\affiliation{\instYamagata}
\affiliation{\instYerevan}
\affiliation{\instYonsei}
  \author{F.~Abudin{\'e}n}\affiliation{\instTriesteINFN} 
  \author{I.~Adachi}\affiliation{\instKEK}\affiliation{\instSOKENDAI} 
  \author{H.~Aihara}\affiliation{\instUTokyo} 
  \author{N.~Akopov}\affiliation{\instYerevan} 
  \author{A.~Aloisio}\affiliation{\instNapoliUNIV}\affiliation{\instNapoliINFN} 
  \author{F.~Ameli}\affiliation{\instRomaINFN} 
  \author{N.~Anh~Ky}\affiliation{\instIOP}\affiliation{\instITAR} 
  \author{D.~M.~Asner}\affiliation{\instBNL} 
  \author{T.~Aushev}\affiliation{\instHSE} 
  \author{V.~Aushev}\affiliation{\instKyiv} 
% ...
\collaboration{Belle II Collaboration}

%% file: acknowledgements.tex
We thank the SuperKEKB group for the excellent operation of the
accelerator; the KEK cryogenics group for the efficient
operation of the solenoid; and the KEK computer group for
on-site computing support.
This work was supported by the following funding sources:
Science Committee of the Republic of Armenia Grant No. 18T-1C180;
Australian Research Council and research grant Nos.
DP180102629, 
DP170102389, 
DP170102204, 
DP150103061, 
FT130100303, 
and
FT130100018; 
Austrian Federal Ministry of Education, Science and Research, and
Austrian Science Fund No. P 31361-N36; 
Natural Sciences and Engineering Research Council of Canada, Compute Canada and CANARIE;
Chinese Academy of Sciences and research grant No. QYZDJ-SSW-SLH011,
National Natural Science Foundation of China and research grant Nos.
11521505,
11575017,
11675166,
11761141009,
11705209,
and
11975076,
LiaoNing Revitalization Talents Program under contract No. XLYC1807135,
Shanghai Municipal Science and Technology Committee under contract No. 19ZR1403000,
Shanghai Pujiang Program under Grant No. 18PJ1401000,
and the CAS Center for Excellence in Particle Physics (CCEPP);
the Ministry of Education, Youth and Sports of the Czech Republic under Contract No.~LTT17020 and 
Charles University grants SVV 260448 and GAUK 404316;
European Research Council, 7th Framework PIEF-GA-2013-622527, 
Horizon 2020 Marie Sklodowska-Curie grant agreement No. 700525 `NIOBE,' 
and
Horizon 2020 Marie Sklodowska-Curie RISE project JENNIFER2 grant agreement No. 822070 (European grants);
L'Institut National de Physique Nucl\'{e}aire et de Physique des Particules (IN2P3) du CNRS (France);
BMBF, DFG, HGF, MPG, AvH Foundation, and Deutsche Forschungsgemeinschaft (DFG) under Germany's Excellence Strategy -- EXC2121 ``Quantum Universe''' -- 390833306 (Germany);
Department of Atomic Energy and Department of Science and Technology (India);
Israel Science Foundation grant No. 2476/17
and
United States-Israel Binational Science Foundation grant No. 2016113;
Istituto Nazionale di Fisica Nucleare and the research grants BELLE2;
Japan Society for the Promotion of Science,  Grant-in-Aid for Scientific Research grant Nos.
16H03968, 
16H03993, 
16H06492,
16K05323, 
17H01133, 
17H05405, 
18K03621, 
18H03710, 
18H05226,
19H00682, 
26220706,
and
26400255,
the National Institute of Informatics, and Science Information NETwork 5 (SINET5), 
and
the Ministry of Education, Culture, Sports, Science, and Technology (MEXT) of Japan;  
National Research Foundation (NRF) of Korea Grant Nos.
2016R1\-D1A1B\-01010135,
2016R1\-D1A1B\-02012900,
2018R1\-A2B\-3003643,
2018R1\-A6A1A\-06024970,
2018R1\-D1A1B\-07047294,
2019K1\-A3A7A\-09033840,
and
2019R1\-I1A3A\-01058933,
Radiation Science Research Institute,
Foreign Large-size Research Facility Application Supporting project,
the Global Science Experimental Data Hub Center of the Korea Institute of Science and Technology Information
and
KREONET/GLORIAD;
Universiti Malaya RU grant, Akademi Sains Malaysia and Ministry of Education Malaysia;
Frontiers of Science Program contracts
FOINS-296,
CB-221329,
CB-236394,
CB-254409,
and
CB-180023, and SEP-CINVESTAV research grant 237 (Mexico);
the Polish Ministry of Science and Higher Education and the National Science Center;
the Ministry of Science and Higher Education of the Russian Federation,
Agreement 14.W03.31.0026;
University of Tabuk research grants
S-1440-0321, S-0256-1438, and S-0280-1439 (Saudi Arabia);
Slovenian Research Agency and research grant Nos.
J1-9124
and
P1-0135; 
Agencia Estatal de Investigacion, Spain grant Nos.
FPA2014-55613-P
and
FPA2017-84445-P,
and
CIDEGENT/2018/020 of Generalitat Valenciana;
Ministry of Science and Technology and research grant Nos.
MOST106-2112-M-002-005-MY3
and
MOST107-2119-M-002-035-MY3, 
and the Ministry of Education (Taiwan);
Thailand Center of Excellence in Physics;
TUBITAK ULAKBIM (Turkey);
Ministry of Education and Science of Ukraine;
the US National Science Foundation and research grant Nos.
PHY-1807007 
and
PHY-1913789, 
and the US Department of Energy and research grant Nos.
DE-AC06-76RLO1830, 
DE-SC0007983, 
DE-SC0009824, 
DE-SC0009973, 
DE-SC0010073, 
DE-SC0010118, 
DE-SC0010504, 
DE-SC0011784, 
DE-SC0012704; 
and
the National Foundation for Science and Technology Development (NAFOSTED) 
of Vietnam under contract No 103.99-2018.45.